\documentclass[12pt]{article}

\usepackage{latexsym,amsmath,amssymb,amsthm,amsfonts,graphicx}
\usepackage{natbib}
\usepackage{epsfig,bm,authblk,url,blkarray}

\usepackage{float} 
\usepackage{booktabs} 
\usepackage{graphicx} 
\usepackage[margin=1cm]{caption} 
\usepackage{mathtools}
\usepackage{xcolor}
\usepackage{color}
\floatstyle{plain}
\usepackage{multirow}
\usepackage[flushleft]{threeparttable}

\usepackage{algcompatible}
\usepackage{algorithmicx}
\usepackage[noend]{algpseudocode}
\usepackage{algorithm}
\usepackage{diagbox}
\usepackage{nicefrac}
\newfloat{Algorithm}{thp}{lop}
\floatname{Algorithm}{Algorithm}

\usepackage{titling}
\newcommand{\m}[1]{\mathrm{#1}}

\begin{document}

\title{Structured variational approximations with skew normal 
decomposable graphical models}
\date{\empty}
\author{Robert Salomone\thanks{\textit{Centre for Data Science, Queensland University of Technology}.}, Xuejun Yu\thanks{\textit{Department of Statistics and Data Science, National University of Singapore}}, David J. Nott\thanks{Corresponding author:  standj@nus.edu.sg. \textit{Department of Statistics and Data Science, National University of Singapore} and \textit{Institute of Operations Research and Analytics, National University of Singapore}.} \, and Robert Kohn\thanks{\textit{UNSW School of Business, School of Economics, University of New South Wales} and 
\textit{Australian Centre of Excellence for Mathematical and Statistical Frontiers (ACEMS)}.}}

\maketitle
\vspace{-0.5in}

\begin{abstract}
Although there is much recent work developing flexible variational methods for Bayesian computation, 
Gaussian approximations with structured covariance matrices are often preferred computationally in high-dimensional settings.  
This paper considers approximate
inference methods for complex latent variable models 
where the posterior is close to Gaussian, but with 
some skewness in the posterior marginals.
We consider skew decomposable graphical 
models (SDGMs), which are based on the closed skew normal family
of distributions, as
variational approximations.  These approximations can
reflect the true posterior conditional independence structure and capture posterior skewness. 
Different parametrizations are explored for this variational family, 
and the speed of convergence and quality of the approximation 
can depend on the parametrization used.  
To increase flexibility, 
implicit copula SDGM approximations  
are also developed, where elementwise transformations of 
an approximately standardized SDGM random vector are considered. 
Our parametrization of the implicit copula approximation 
is novel, even in the special case of a Gaussian approximation.
Performance of the methods is examined in a number of real examples involving generalized linear
mixed models and state space models, and
we conclude that our copula approaches are most accurate, 
but that the SDGM methods are often nearly as good and have lower
computational demands.


\end{abstract}

\section{Introduction}\label{sec:Intro}

Variational inference \citep{ormerod2010explaining,blei2017} is an attractive scalable alternative
to conventional methods for Bayesian computation.  Variational methods optimize an approximation
to a Bayesian posterior distribution within some chosen family.  
In choosing a suitable form for the approximation, it is important to balance flexibility 
and computational tractability.  The kind of flexibility required depends on the problem at hand, 
and here we consider high-dimensional problems with some known conditional independence
structure in the target posterior distribution.  We use approximations which preserve the conditional
independence structure, and we are particularly interested in
approximating posterior distributions for latent variable models, such as random effects models and state space models.

Our paper makes three contributions.  First, we extend Gaussian graphical models
and use more flexible skew decomposable graphical models (SGDMs)
\citep{zareifard+rkl16} for variational inference, 
to allow skewed marginal distributions in our
approximations.  SDGMs are based on the closed skew normal family
of distributions, and conditional independence structure 
is imposed through sparsity in the precision matrix, 
which reduces the number of variational parameters to
optimize in high-dimensional settings.
Second, we explore different parametrizations of the variational family, and show that this can be important for simplifying the optimization and
obtaining better quality approximations.  
Third, we make the 
approximations more flexible by  
transforming each marginal and then fitting an SGDM to the transformed marginals. We call this an SDGM implicit copula variational family \citep{smith23}, 
and consider 
the sinh-arcsinh elementwise transformations, which
have not been used previously for related Gaussian copula approximations.    
Even for the Gaussian case, the parametrization of our implicit
copula approximation is novel.  
We conclude that both the SDGM and SDGM copula approximations
can be effective for capturing skewed marginal distributions
in latent variable models.  However, the copula
methods are superior overall in terms of the quality of
the approximation, while the SDGM methods can
perform nearly as well with reduced computational demands.

Developing highly flexible approximations for variational inference is a focus of recent variational
inference research.  Approaches to this problem include normalizing flows \citep{rezende+m15}, mixture models \citep{Jaakkola1998,salimans2013fixed,Guo2016,Miller2016,jerfel2021}, 
and copulas \citep{han2016,tran+ba15,smith+ln20,smith+l21,gunawan+kn21b}, among others.  
In high-dimensional problems, it is useful to consider modest extensions of Gaussian approximations
incorporating some ability to capture posterior skewness.  One possibility is to
use a multivariate skew normal family, which is first considered in \cite{ormerod11}, using
one-dimensional quadrature methods for performing the variational optimization.  
Natural gradient optimization methods for skew normal families are discussed in \cite{lin+ks19}, 
and implicit copulas of skew-normal densities are considered by \cite{smith+ln20}, where the authors
consider a factor structure for covariance matrices. 
\cite{fasano+dz22} consider variational approximations for high-dimensional probit regression, 
and their partially factorized approximation belongs to the class of unified skew normal densities.  These previous uses of skew normal variational approximations
do not attempt to match any conditional independence structure in the true
posterior distribution in a general setting, which is the focus of the present work.

One approach to developing flexible structured variational inference methods with
conditional independence structure is to generalize Gaussian
approximations having sparsity in the precision matrix such as those of \cite{Archer2016} and \cite{tan2017}.  
For example, \cite{tan+bn20} consider a sequential decomposition of the posterior distribution into
a marginal distribution for global variables and conditional distribution for local latent variables given global ones, 
with each term in the decomposition being a Gaussian density.  The marginal distribution of local latent variables
can be non-Gaussian in their approach. 
\cite{tan21} considers a reparametrized variational Bayes (VB) approach, where the reparametrization of the local latent variables
depends on the global variable.  This leads to a non-Gaussian
approximation in the original parametrization with greatly 
improved accuracy.  
\cite{quiroz+nk18} combines elements of factor structure and conditional independence structure through
sparsity of the precision matrix to obtain Gaussian approximations suitable for high-dimensional state space models.
Another approach is structured stochastic variational inference \citep{hoffman+b15}, 
which applies in models with conjugate structure.  This 
generalizes variational inference methods for latent variable models 
in \cite{hoffman2013} to the setting of non-factorized approximations.   \cite{lucia-etal21}
consider an automated stochastic variational inference approach where approximations follow the parameteric form of the prior.  \cite{ambrogioni+sv21} consider
a type of normalizing flow (cascading flows) which is able
to respect graphical structure.  
\cite{nolan+mw20} consider mean field and variational message passing algorithms for 
regression models with higher level random effects.
\cite{agrawal+d21} consider Gaussian approximations with
amortized inference for local latent variables for large-scale
applications.

Variational approximations for complex latent variable models can also be formed by combining elements
of variational inference and Monte Carlo methods such as MCMC. 
\cite{ruiz+t19} consider choosing an initial parametrized distribution, which is then
updated using a small number of MCMC steps.  The parameters in the initial distribution 
interact with the MCMC kernel used in the variational optimization.  \cite{loaiza+snd22} 
consider a method in which a parametric variational
family for some of the model parameters is combined with
the exact conditional posterior distribution for the rest.  Reparametrization gradients for optimization
can be obtained where a few steps of MCMC are used for 
sampling the parameters which follow the exact posterior conditional in the approximation.  Related approaches
were earlier considered in \cite{gunawan+tk17}, where the authors focus on random effects models and use importance sampling
rather than MCMC, and by \cite{hoffman17} who consider maximum likelihood estimation in latent variable models.
Application of the approach of \cite{loaiza+snd22} to stochastic volatility models is considered
in \cite{gunawan+kn21}, where they also combine the approach with the methods of \cite{tan+bn20} and \cite{smith+ln20}.  
\cite{gopelrud22} considers mean field approximations for binary random effects models with arbitrarily many levels
using data augmentation and a post-processing adjustment involving an MCMC step.  
\cite{naesseth+lb20} consider the use of MCMC in a Markovian
score climbing algorithm for minimizing the inclusive
Kullback-Leibler divergence.
There are a variety
of other methods combining MCMC or sequential Monte Carlo and variational inference, and our review of the literature here
is not intended to be comprehensive.  

An alternative approximate inference method to variational approximation
is integrated nested Laplace approximation (INLA) \citep{rue+mc09} 
which is used
for latent Gaussian models.  When  applicable, 
the INLA methodology is faster due to exploiting the 
assumed latent Gaussian structure, and \cite{chiuchiolo+vr22} considers some variants of the method which are 
particularly effective when skewed approximations are needed.   However, variational methods can be used
for a wider class of models than INLA.

The next section gives some background on variational inference methods and describes the 
SDGM family of approximations that we use in our work.  
Section 3 describes our approach to optimizing the approximation, considering different parametrizations
of the variational family and also extensions including sinh-arcsinh
marginal transformations.  
Section 4 compares the methods considered 
in several real examples, and Section 5 concludes.
The paper also has an online supplement that presents extra simulation results. 

\section{SDGM variational approximations}\label{sec:variational}

\subsection{Variational inference}

Let $p(y|\theta)$ be the likelihood for parameter $\theta$ with
$p(\theta)$ its prior;  the posterior density, given the data $y$,
is $p(\theta|y)\propto p(\theta)p(y|\theta)\coloneqq h(\theta)$.
Variational inference methods perform Bayesian computation by optimizing a measure of closeness between
the posterior density $p(\theta|y)$ and an approximation $q_\lambda(\theta)$, where $\lambda$ are 
variational parameters to be optimized.  For example, if $q_\lambda(\theta)$ is multivariate Gaussian, 
$\lambda$ may be the mean vector and covariance matrix.
The Kullback-Leibler divergence is usually the measure of
closeness that is optimized, 
\begin{align}
\text{KL}(q_\lambda(\theta)|| p(\theta|y)) & \coloneqq \int q_\lambda(\theta) \log \frac{p(\theta|y)}{q_\lambda(\theta)}\,d\theta, \label{KLD}
\end{align}
and minimizing (\ref{KLD}) with respect to $\lambda$ is equivalent to maximizing the evidence lower bound (ELBO), defined
as
\begin{align}
{\cal L}(\lambda) & \coloneqq \int \log \frac{h(\theta)}{q_\lambda(\theta)} q_\lambda(\theta)\,d\theta.  \label{elbo}
\end{align}
For models with conjugate structure, and using factorized posterior approximations, it is often possible to perform
the optimization using a coordinate ascent scheme with closed form updates 
(see, for example, \cite{ormerod2010explaining} and \cite{blei2017}).  We use stochastic gradient ascent
methods for the optimisation as they are easier
to implement for many models of interest.

\subsection{SDGM family}

We consider a variational
approximation taking the form of a skew decomposable
graphical model (SDGM) \citep{zareifard+rkl16}.  If $p$ is
the dimension of $\theta$, the SDGM variational
approximation is parametrized by a location vector $\mu\in \mathbb{R}^p$, 
a vector of skewness parameters $\alpha\in \mathbb{R}^p$, 
a lower triangular matrix $L$ with ones on the diagonal, 
and a vector $\kappa\in \mathbb{R}^p$ with positive entries.   
A precision matrix $Q$ is defined from $\kappa$ and $L$ through
a modified Cholesky decomposition, $Q=L D_\kappa^2 L^\top$, 
$D_\kappa=\text{diag}(\kappa)$, where $\text{diag}(a)$ for vector $a$ denotes the diagonal matrix with diagonal
entries $a$.  Below we write $a\odot b$ for the elementwise
product of two vectors $a$ and $b$, and define
$D_\alpha=\text{diag}(\alpha)$.  Our notation is similar
to \cite{zareifard+rkl16}, but they define $L$ as an upper
triangular matrix, whereas here it is defined to be lower triangular.  They also define $D_\kappa=\text{diag}(\kappa\odot \kappa)$, whereas we define it as $D_\kappa=\text{diag}(\kappa)$.
SDGMs belong to the closed
 skew normal family of distributions \citep{gonzalez-farias+dg04}, 
 and this gives them more convenient properties than 
 previously proposed graphical
 models for multivariate skew normal densities \citep{capitanio+as03}. 
   
The lower triangular matrix $L$ in the SDGM is 
typically sparse, with the pattern of zeros relating to the conditional independence
 structure of the distribution, which is explained further below.
The SDGM variational approximation of $p(\theta|y)$ has the density
\begin{align}
  q_\lambda(\theta) & = 2^p \phi(\theta;\mu, Q^{-1}) \prod_{k=1}^p \Phi\left(\left\{D_\kappa D_\alpha L^\top (\theta-\mu)\right\}_k\right),  \label{sng-density}
 \end{align}
 where $\phi(x;\nu,\Sigma)$ denotes the multivariate normal density with mean vector $\nu$ and covariance matrix $\Sigma$, $\Phi(\cdot)$ denotes the univariate standard normal distribution function and $\left\{D_\kappa D_\alpha L(\theta-\mu)\right\}_k$ denotes the $k$th element of $D_\kappa D_\alpha L(\theta-\mu)$.  Here, $\lambda$ denotes the set of variational parameters $\lambda=(\mu^\top,\alpha^\top,\kappa^\top,\text{vech}(L)^\top)^\top$, where  $\text{vec}(\cdot)$ is the vectorization operator that stacks
the elements of a matrix into a vector proceeding columnwise from left to right, and $\text{vech}(\cdot)$ is the half vectorization operator that 
stacks the elements of the lower triangle of a square matrix.  

\citet{zareifard+rkl16} note that if $\theta\sim q_\lambda(\theta)$, then this is equivalent to 
 \begin{align}
 \theta & = \mu+L^{-\top}\left(\alpha \odot \kappa^{-1}\odot ({\bf 1}+\alpha^2)^{-1/2}\odot |U|+\kappa^{-1}({\bf 1}+\alpha^2)^{-1/2} \odot V\right), \label{genrep}
 \end{align}
where taking absolute values and powers is defined elementwise for vectors,  
$U,V\sim  N(0,\m I_p)$ and ${\bf 1}$ denotes a $p$-dimensional vector of ones.
This expression further simplifies to
 \begin{align*}
 \theta & = \mu+L^{-\top}\left(\kappa^{-1} \odot ({\bf 1}+\alpha^2)^{-1/2} \odot \left\{\alpha \odot|U|+ V\right\}\right).
  \end{align*}
The above expressions are important later for obtaining low variance gradient estimates in the stochastic
gradient optimization of the ELBO.

The sparsity pattern of the matrix $L$ in the SDGM is defined from a graph ${\cal G}$ encoding the conditional
independence properties of the distribution.  
In the SDGM, $\theta_i$ and $\theta_j$ are conditionally independent
given the remaining variables if  $Q_{ij}=0$, and so conditional independence structure is determined by
the sparsity structure of the precision matrix $Q$.  In the SDGM,  the conditional independence structure is described by 
a decomposable graph; it is unnecessary to give a precise definition of this here, and we refer the reader to
\cite{zareifard+rkl16} for further discussion or 
\cite{lauritzen96} for a textbook introduction to graphical models.
For a decomposable graph, there is an ordering of the variables
such that the sparsity structure of the lower triangle of $L$ reflects that of $Q$.  That is, 
if $i>j$, $\theta_i$ and $\theta_j$ are conditionally independent in the SDGM given the remaining
variables if $L_{ij}=0$.  
Our paper focuses on approximating the posterior distribution for latent variable models such as
longitudinal random effects models and state space models, 
and in these models the 
conditional independence structure of the posterior distribution 
can be expressed in terms of a decomposable graph.    
These models have global parameters denoted as $\eta$, and local
latent variables denoted as $b_1,\dots, b_n$.  For example, in a longitudinal random effects model,  $\eta$ contains fixed
effects and variance parameters, and $b_i$ is the random effect for observation $i$.  In a state space model, the
local latent variables correspond to the states at different times.  Write $\theta=(b_1^\top ,\dots, b_n^\top ,\eta^\top )^\top$
for the set of unknowns in the model.  

Following \cite{tan2017}, consider a model where the likelihood is
$$\prod_{i=1}^n p(y_i|\theta,b_i);$$
 $n$ is the number of observations, and the prior is
$$p(\theta)p(b_1,\dots, b_k|\theta)\prod_{i=k+1}^{n} p(b_i|b_{i-1},\dots, b_{i-k},\theta).$$
This model is general enough to include both random effects models ($k=0$) and state space models ($k=1$) as
special cases;  the model has conditionally independent observations given the global parameters and local
latent variables, and the prior on the latent variables is Markovian of order $k$, where $Q$ 
and $L$ are partitioned into blocks
conformably with $(b_1^\top,\dots, b_n^\top,\eta^\top)^\top$. For the case of a random effects model,  the appropriate structure for 
$Q$ is
$$
Q  = \left[\begin{array}{ccccc}
  \bar{Q}_{11} & 0 & \ldots & 0 & \bar{Q}_{1,n+1} \cr
 0 & \bar{Q}_{22} & \ldots & 0 & \bar{Q}_{2,n+1} \cr
 \vdots & \vdots & \ddots & \vdots & \vdots \cr
 0 &  0 & \ldots & \bar{Q}_{nn} & \bar{Q}_{n,n+1} \cr
 \bar{Q}_{n+1,1} & \bar{Q}_{n+1,2} & \ldots & \bar{Q}_{n+1,n} & \bar{Q}_{n+1,n+1} 
 \end{array}\right],
$$
where we write $\bar{Q}_{ij}$ for the $(i,j)$th block entry.  
For a state space model, 
$$
Q = \left[\begin{array}{ccccccc}
\bar{Q}_{11} & \bar{Q}_{21}^\top  & 0 & \ldots & 0 & 0 & \bar{Q}_{n+1,1}^\top \cr
\bar{Q}_{21} & \bar{Q}_{22} & \bar{Q}_{32}^\top  & \ldots & 0 & 0 & \bar{Q}_{n+1,2}^\top \cr
0 & \bar{Q}_{32} & \bar{Q}_{33}  & \ldots & 0 & 0 &  \bar{Q}_{n+1,3}^\top \cr
\vdots & \vdots & \vdots & \ddots & \vdots & \vdots & \vdots \cr
0 & 0 & 0  &  \ldots  & \bar{Q}_{n-1,n-1} & \bar{Q}_{n,n-1}^\top & \bar{Q}_{n+1,n-1}^\top \cr
0 &  0 & 0  & \ldots & \bar{Q}_{n,n-1} & \bar{Q}_{nn} & \bar{Q}_{n+1,n}^\top \cr
\bar{Q}_{n+1,1} & \bar{Q}_{n+1,2} & \bar{Q}_{n+1,3} & \ldots & \bar{Q}_{n+1,n-1} & \bar{Q}_{n+1,n} & \bar{Q}_{n+1,n+1} 
\end{array}\right].$$
It is easy to see (\cite{rothman+lz10}, Proposition 1) that the 
block sparse structure of the lower triangle of $L$ follows that of $Q$ in both cases above.

\section{Optimizing the SDGM approximation}\label{sec:optimizing}

Optimizing  the value of $\lambda$ in \eqref{sng-density} so that $q_\lambda(\theta)$ is
closest to the posterior density $p(\theta|y)$ in the Kullback-Leibler sense is
equivalent to optimizing the ELBO (\ref{elbo}). 
The optimization is done by stochastic gradient ascent, where starting  
 from some initial value $\lambda^{(0)}$ for the variational parameters we update by
 $$\lambda^{(t+1)}=\lambda^{(t)}+\delta_t\odot \widehat{\nabla_\lambda {\cal L}(\lambda^{(t)})},$$
for $t\geq 0$ until some stopping rule is satisfied;  here $\delta_t$ is a vector of step sizes of the same dimension
as $\lambda$ and $\widehat{\nabla_\lambda {\cal L}(\lambda^{(t)})}$ is an unbiased estimate
of $\nabla_\lambda {\cal L}(\lambda^{(t)})$.  The choice
of these learning rates in our examples is discussed later.

For stable and fast optimization 
convergence it is 
important to have low variance unbiased gradient estimates.
The generative representation (\ref{genrep}) is the basis for application of the 
so-called ``reparametrization trick'' \citep{Kingma+w13,rezende+mw14} for
variance reduction in unbiased estimation of the ELBO gradients.  
Considering the generative representation (\ref{genrep}) with
$\theta=\theta(U,V,\lambda)$, $\nabla_\lambda {\cal L}(\lambda)$ can be written as
\citep{han2016,Roeder2017}
 \begin{align}
\nabla_\lambda {\cal L}(\lambda) & = \int \frac{d\theta}{d\lambda}^\top \left\{ \nabla_\theta \log h(\theta)-\nabla_\theta \log q_\lambda(\theta)\right\}\, \phi(u) \phi(v)\,du\,dv. \label{repar-grad}
 \end{align}
Equation (\ref{repar-grad}) is an expectation with respect to the standard Gaussian density of $(U,V)$, and can be 
estimated unbiasedly by one or more Monte Carlo samples.  Appendix A gives details of reparametrization gradients 
for the SDGM approximating family.  Computation of the gradient estimates is done by efficiently solving sparse triangular linear systems involving $L$.  The examples
later consider an alternative implementation via
 automatic differentiation capabilities using
\verb|PyTorch| \citep{paszke2017}; this is also
discussed in the appendices.

\subsection{An alternative parametrization}

In statistical inference for variants of the multivariate 
skew normal distribution it is well-known that likelihood-based inference
can be difficult in the usual direct parametrization of such
distributions.  Singularity of the Fisher information
can occur when $\alpha=0$, and  
this can be avoided by various ``centered'' parametrizations \citep{arrelano-valle+a08}.  We now show that these reparametrizations are also useful for our SDGM 
variational approximation.  The centred parametrization discussed next is also
important in constructing more flexible copula approximations
in the next subsection.

We rewrite equation (\ref{genrep}) as 
\begin{align}
 \theta & =  \mu+L^{-\top} (\kappa^{-1}\odot Z_\alpha),  \label{genrep2}
\end{align}
where 
$$Z_\alpha=\alpha\odot ({\bf 1}+\alpha^2)^{-1/2}\odot |U|+({\bf 1}+\alpha^2)^{-1/2}\odot V),$$
with the $k$th component of $Z_\alpha$ is skew normal, $SN(0,1,\alpha_k)$.  
Define $\delta\coloneqq \alpha\odot (1+\alpha^2)^{-1/2}$, $\mu(\alpha)\coloneqq \delta\odot \sqrt{2/\pi}$
and 
$$\sigma(\alpha)\coloneqq\left(1-\frac{2\delta^2}{\pi}\right)^{1/2}.$$
The vectors $\mu(\alpha)$ and $\sigma(\alpha)$ contain the means and standard deviations
of the components of $Z_\alpha$ respectively.  Next, define a centered version of $Z_\alpha$ having components
with mean zero and variance one,
$$Z_\alpha^c\coloneqq (Z_\alpha-\mu(\alpha))\oslash \sigma(\alpha),$$
where for $p$-vectors $a$ and $b$ we write $a\oslash b$ for the vector with $i$th entry $a_i/b_i$, $i=1,\dots, p$,
provided all entries of $b$ are nonzero.
Then, $Z_\alpha=\mu(\alpha)+\sigma(\alpha)\odot Z_\alpha^c$, and plugging this expression into (\ref{genrep2}) 
we obtain
\begin{align}
  \theta & = \xi+L^{-\top} (\nu\odot Z_\alpha^c),  \label{cengenrep}
\end{align}
where $\xi=\mu+L^{-\top} D_\kappa^{-1} \mu(\alpha)$ and $\nu=\kappa^{-1}\odot \sigma(\alpha)$.  

We now consider a new parametrization of the SDGM variational approximation, where instead of using
the parameters $\lambda=(\mu^\top,\alpha^\top,\kappa^\top,\text{vec}(L)^\top)^\top$, we use
$\rho=(\xi^\top,\alpha^\top,\nu^\top,\text{vec}(L)^\top)^\top$.  In the original parametrization, 
the mean of the variational distribution is a function of all the variational parameters, whereas after reparametrization
the mean is $\xi$.  Similarly, after reparametrization the vector of component standard deviations is only
a function of $\nu$ and ${L}$, whereas previously this was a function of $\alpha$, $\kappa$ and ${L}$. 
The reparametrization simplifies the dependence between the parameters in the variational optimization.
Write $q_\rho(\theta)$ for the variational approximation in the new parametrization.  Appendix B details 
reparametrization gradients for the centered parametrization.  These
computations can again be done efficiently using solutions of sparse triangular linear systems involving $L$.  

\subsection{SDGM implicit copula with sinh-arcsinh marginal transformations}

We now consider making the SDGM approximations more flexible by
considering marginal transformations of an SDGM random vector,
giving an implicit SDGM copula approximating family.  See \cite{han2016}, \cite{smith+ln20} and \cite{smith+l21} for further 
discussion of implicit copula variational approximations.  Write 
$t_g(z):\mathbb{R}\rightarrow \mathbb{R}$, $g\in G$, for a 
family of one-to-one transformations, where 
$g$ is a parameter that can be chosen.  
We consider variational approximations obtained by transforming
an approximately standardized SDGM random vector using 
$t_g(z)$ elementwise, 
where $g$ varies across components, 
and then adding a location and scale adjustment.  
Later we use the inverse of the sinh-arcsinh transformation 
\citep{jones+p09} for $t_g(z)$, 
\begin{align}
  t_g(z) & \coloneqq \text{sinh}\left\{\delta^{-1}\odot  \left\{\text{sinh}^{-1}(z)+\epsilon\right\}\right\}, \label{inv-sas-transform}
\end{align}
where $g=(\epsilon,\delta)^\top$, with $\epsilon \in \mathbb{R}$ a skewness
parameter and $\delta>0$ a kurtosis parameter.  The sinh-arcsinh
transformation $t_g^{-1}(\cdot)$ is
\begin{align}
  t_g^{-1}(z) & = \text{sinh}\left\{\delta \odot  \text{sinh}^{-1}(z)-\epsilon\right\}. \label{sas-transform}
\end{align}
If $Z$ is standard normal, the random variable $t_g(Z)$ is positively (negatively)
skewed if $\epsilon>0$ ($\epsilon<0$), and has heavier (lighter) tails
than normal if $\delta<1$ ($\delta>1$);   $\epsilon=0$ and $\delta=1$ is the identity transformation.  

We consider a variational approximation corresponding to
the generative model
\begin{align}
  \theta & = \xi+\exp(\bar{\nu})\odot  t_\gamma(L^{-\top}Z_{\alpha}^c), \label{gen-sas}
\end{align}
where $Z_\alpha^c$ and $\alpha$ are 
defined in section 3.1 for the centred parametrization of the SDGM
approximation, $\xi$ is a vector of location parameters, 
$L$ is a lower-triangular
matrix with diagonal elements $1$, $\bar{\nu}=\log \nu$, where $\nu$
is defined in Section 3.1 and the log is taken elementwise, and
for $w\in \mathbb{R}^p$,
$$t_\gamma(w)=(t_{\gamma_1}(w_1),\dots, t_{\gamma_p}(w_p))^\top,$$
with $\gamma=(\gamma_1,\dots, \gamma_p)^\top$ being a vector
of marginal transformation parameters.  In \eqref{gen-sas}
the vector $L^{-\top}Z_\alpha^c$ is transformed nonlinearly by 
$t_\gamma(\cdot)$. Note that
$L^{-\top}Z_\alpha^c$ has zero mean (since $Z_\alpha^c$ has
zero mean) and it is on a roughly standardized scale, since elements of
$Z_\alpha^c$ have standard deviation $1$ and $L$ has unit
diagonal.  

\cite{smith+l21} discuss the importance of using a centred and standardized
random vector in constructing implicit copula variational approximations.  They consider implicit elliptical copulas 
where a mean and scale shift are applied only after elementwise
nonlinear transformations of a standardized random vector are made.  
The motivation for their approach is that the previous implicit
Gaussian and skew Gaussian copula approximations of \cite{smith+ln20}
are not invariant to location shifts.   Using a spherical 
factor parametrization
of a correlation matrix for the copula, they construct approximations
that do possess a location invariance property, and show that this
results in higher quality approximations.  
Our SDGM approximations 
do not use a factor structure for the covariance matrix because we
wish to capture the conditional independence structure of the true
posterior, and hence we
cannot use the reparametrization of \cite{smith+l21}.  However,
the generative model \eqref{gen-sas} where the transformation 
$t_\gamma(z)$ is applied to $L^{-\top}Z_\alpha^c$ 
achieves a similar goal, starting from the centred parametrization of section 3.1.  

It may not be immediately obvious that \eqref{gen-sas} is equivalent
to using the centered parametrization of Section 3.1 when 
$t_\gamma(\cdot)$ is the identity transformation.  To understand how 
\eqref{gen-sas} is obtained in this case, write \eqref{cengenrep} as \\
$$\theta=\xi+D_\nu \left\{D_\nu^{-1}L D_\nu\right\}^{-\top} Z_\alpha^c,$$
where $D_\nu=\text{diag}(\nu_1,\dots, \nu_p)^\top$.  Observe that
$D_\nu^{-1} LD_\nu$ is lower triangular with diagonal elements
$1$ and the same zero entries as $L$.  By overloading notation and writing $L$ instead of $D_\nu^{-1}LD_\nu$, we get
$$\theta=\xi+D_\nu L^{-\top} Z_\alpha^c,$$
which is \eqref{gen-sas} when $t_\gamma(\cdot)$ is the identity transformation.  

To obtain reparametrization gradients for use in stochastic optimization
we need the density of the variational
approximation given by the generative model \eqref{gen-sas}.  First, 
consider $\widetilde{\theta}=L^{-\top}Z_\alpha^c$.  
Recall that
$Z_\alpha^c=(Z_\alpha-\mu(\alpha))\oslash \sigma(\alpha)$, where
$Z_\alpha$ is a vector of independent skew normal random variables, 
$Z_{\alpha,k}\sim SN(0,1,\alpha_k)$.  Then $\widetilde{\theta}$ is 
an SDGM random vector, with parameters $\mu$, $L$, $\alpha$, $\kappa$, with $\mu$ and $\kappa$ functions of $\alpha$ and $L$ as
$\mu=-L^{-\top}\mu(\alpha)\oslash \sigma(\alpha)$ and $\kappa=\sigma(\alpha)$. 
It is straightforward to obtain the SDGM density for
$L^{-\top}Z_\alpha^c$.  A change of variables from 
$\widetilde{\theta}$ to $\theta$ via the elementwise transformation
$\theta=\xi+\exp(\bar{\nu})\odot t_\gamma(\widetilde{\theta})$, results in a 
(diagonal) Jacobian for obtaining the density of $\theta$, which we write as
$q_{\breve{\lambda}}(\theta)$, where $\breve{\lambda}$ consists of the
variational parameters $(\xi,\bar{\nu},\alpha,L,\gamma)$.   

Write $h_g(z')=t_g^{-1}(z')$ for the inverse of $t_g(z)$, and
$h_\gamma(w)=(h_{\gamma_1}(w_1),\dots, h_{\gamma_p}(w_p))^\top,$
for $w\in\mathbb{R}^p$.  Then
$$\widetilde{\theta}=h_\gamma\left((\theta-\xi)\oslash \exp(\bar{\nu})\right).$$
Writing $q_\lambda(\widetilde{\theta})$ for
the SDGM density of $\widetilde{\theta}$, the density of 
$\theta$ is 
\begin{align}
  q_{\breve{\lambda}}(\theta) & =q_\lambda(\widetilde{\theta}) \prod_{j=1}^p \frac{d\widetilde{\theta}_j}{d\theta_j},  \label{qsas}
\end{align}
with
\begin{align}
  \frac{d\widetilde{\theta}_j}{d\theta_j} & =h_{\gamma_j}'\left(\frac{\theta_j-\xi_j}{\exp(\bar{\nu}_j)}\right)\times \frac{1}{\exp(\bar{\nu}_j)}. \label{dthetajtildedthetaj}
\end{align}
Appendix C gives details of the  reparametrization gradients for this variational family.

\section{Examples}\label{sec:example}

We now compare our approximations with other benchmarks in
three examples.  The methods we compare are:
\begin{enumerate}
\item GVA - Gaussian variational approximation, which is the SDGM approximation with $\alpha=0$.  
\item SDGM - The SDGM variational approximation using the direct parametrization.  
\item SDGM-C - The SDGM variational approximation with the centered parametrization of section 3.1.  
\item SDGM+SAS - implicit copula approximation with inverse sinh-arcsinh transformation.
\item GVA+SAS - implicit Gaussian copula approximation with inverse sinh-arcsinh transformation, the SDGM+SAS method with $\alpha=0$. This implicit Gaussian copula approximation uses a novel parametrization
compared to previous Gaussian copula approximations, building
on the centred parametrization of the SDGM model.
\end{enumerate}
Our examples consider three longitudinal random effects models and a state space model. For the three random effects models, two have binary response and one a count response, and both normally distributed and $t$-distributed random effects are considered.

In implementing our variational approaches 
we use a learning rate annealing strategy during training. 
The learning rate is set to a large value for the first 10 or 20 
thousand iterations, and then reduced every 
10 or 20 thousand iterations. This strategy helps to explore 
the space and reach a higher ELBO value.  The MCMC benchmarks
reported are obtained using the \verb+rstan+ software \citep{stan}
using 50,000 iterations, discarding the first 25,000 iterations as burn-in.  
Python code for reproducing the examples
is at \url{https://github.com/Yu-Xuejun}.

\subsection{Six cities data}

The first example is the six cities data \citep{fitzmaurice+l93}, 
from a longitudinal
study of health effects of air pollution.  There are data on 537 children, 
followed annually from ages 7 to 10.  The response $y$ is a binary indicator
for wheezing status ($1$ for yes, $0$ for no).  Write $y_{ij}$ for the $j$th
observation on the $i$th subject $i=1,\dots, n$, $j=1,\dots, 4$. 

A random
intercept logistic regression model 
\begin{align*}
  \text{logit}(p_{ij})=x_{ij}^\top \beta + b_i
\end{align*}
 is fitted, where $p_{ij}$ is the mean of $y_{ij}$, $x_{ij}$ are covariates with fixed effects
$\beta$, and $b_i$ is a random intercept.  Two priors are considered
for $b_i$.  The first is $N(0,\exp(2\zeta))$, where an
$N(0,100)$ hyperprior is used for $\zeta$.  The second
is $t_{10}(0,\exp(2\zeta))$ with the same hyperprior
for $\zeta$.  The prior on $\beta$ is $N(0,100 I)$.  The vector 
$x_{ij}$ is $3\times 1$, consisting of mother's smoking status (\texttt{Smoke}, $1=$yes and $0=$no), age of the child (\texttt{Age}, centred) and an interaction
term (\texttt{Smoke}$\times$\texttt{Age}).  

\begin{figure}[htp]
\begin{center}
\begin{tabular}{c}
\includegraphics[height=90mm]{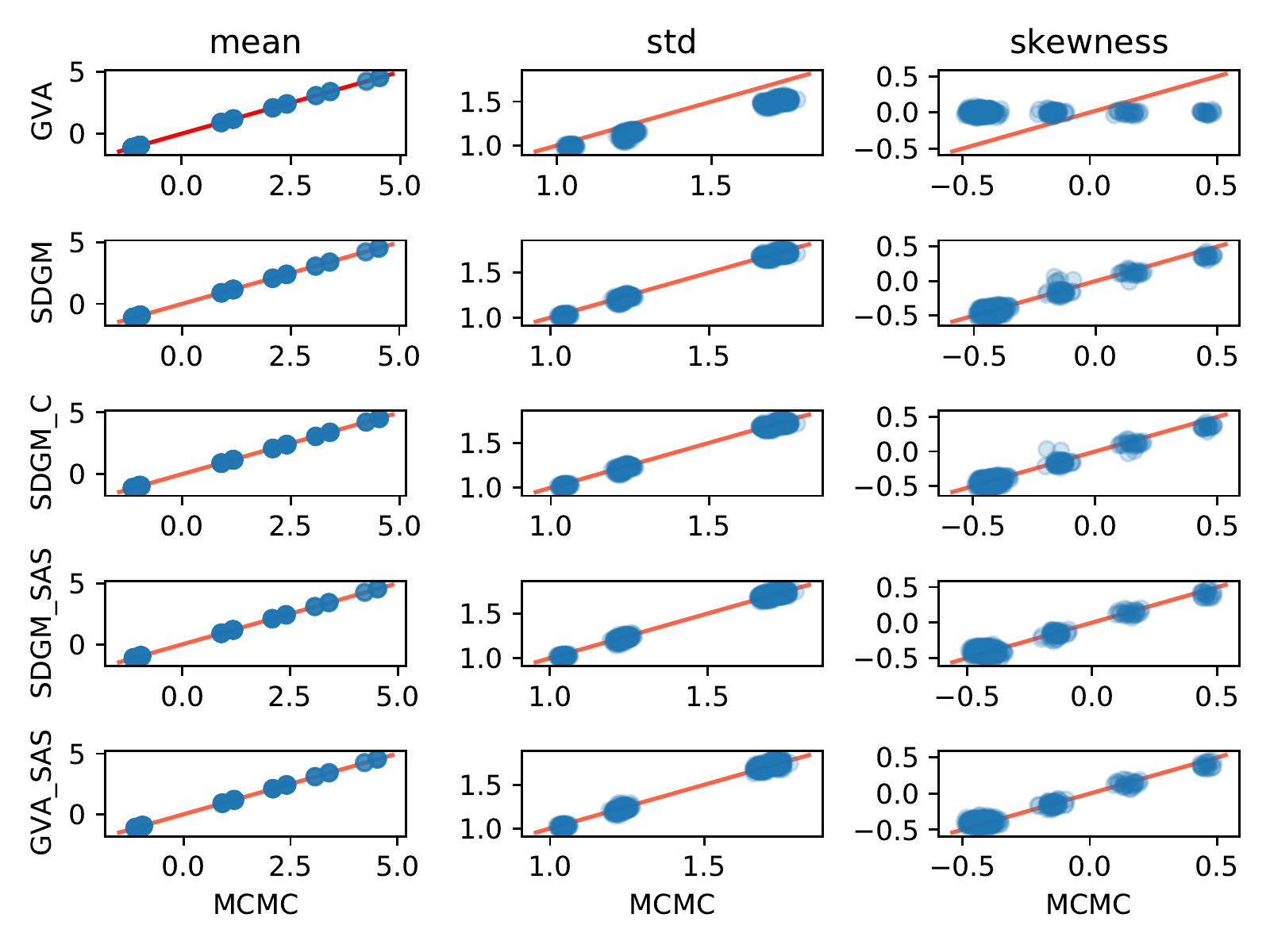} \\
\includegraphics[height=90mm]{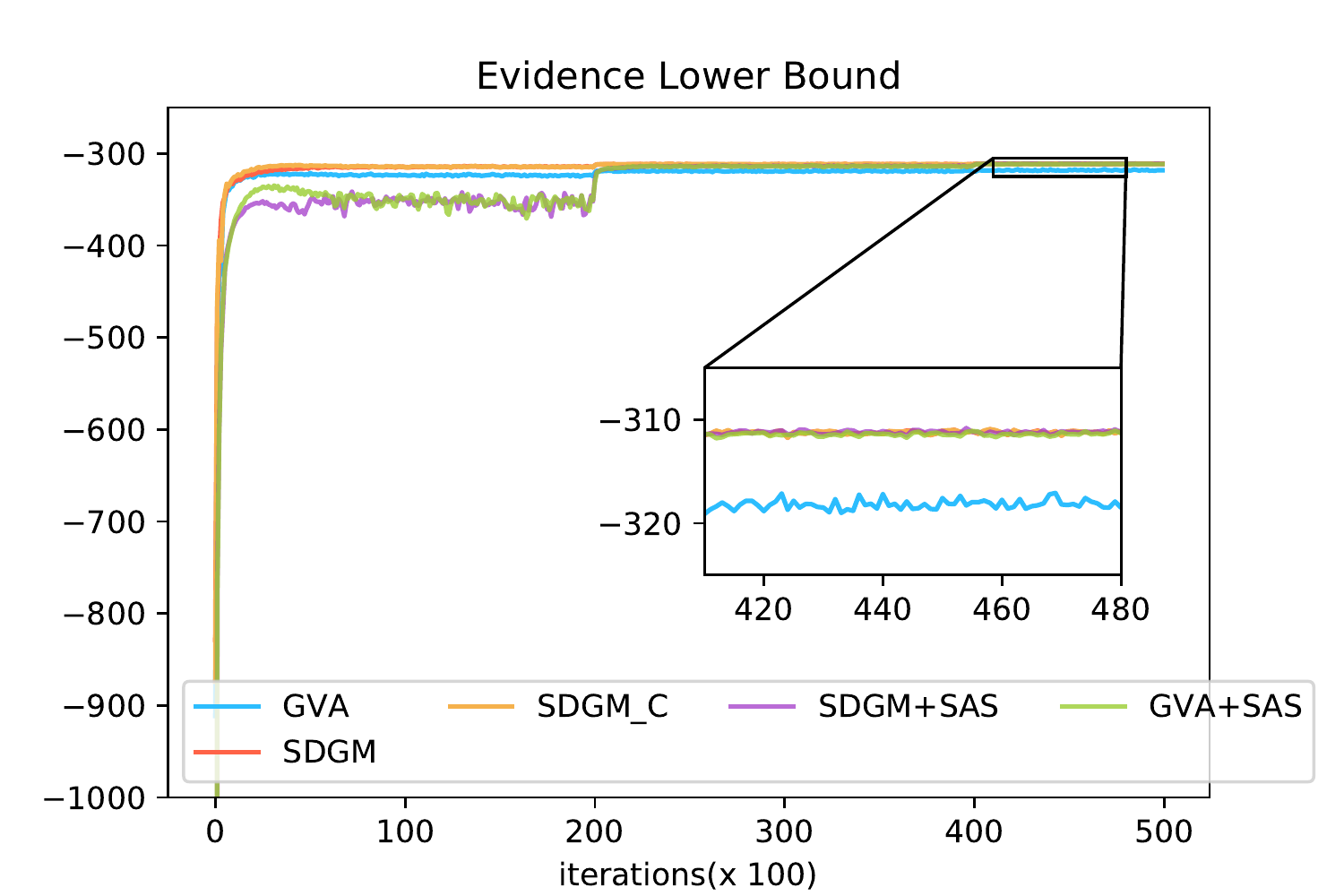} \\
\end{tabular}
\end{center}
\caption{\label{six-cities-normal} 
Comparing the means, standard deviations and skewness for the random effects estimated by MCMC and approximate methods (top) and Monte Carlo estimate of ELBO versus iteration number (bottom) for six cities data and normal random effects.  }
\end{figure}

The top panel of 
Figure \ref{six-cities-normal} considers the quality of the variational
estimates of mean, standard deviation and skewness for
the random effects compared to an MCMC benchmark for the case
of normal random effects.  In the plots, an accurate
approximation is indicated by the points following a diagonal 
line.  
The bottom panel shows the Monte Carlo estimate of the
ELBO versus iteration number.  Appendix D shows
a similar figure for the case of $t$-distributed 
random effects, as well as some plots of marginal
posterior densities of the fixed effects and variance parameters.

We make three observations.  First, for both normal and $t$-distributed random effects, all the SDGM and copula methods are clearly
superior to the Gaussian approximation in terms of the ELBO, as well as
estimating the random effect standard deviations. Second, the two copula
methods are slightly better than SDGM and SDGM-C 
for estimating both standard
deviation and skewness of the random effects when
they are $t$-distributed.  
Third, during our experiments we found that optimizing 
the GVA+SAS approximation is easier in the sense that
different methods for adaptively determining the learning
rates lead to similar solutions, whereas this is not
always the case for the SDGM and SDGM-C approximations.

\subsection{Polypharmacy data}

The polypharmacy dataset \citep{hosmer2013} considers a logistic random effects
model with random intercept for binary responses $y_{ij}$, $i=1,\dots, 500$, 
$j=1,\dots, 7$ where $y_{ij}=1$ if subject $i$ in year $j$ of the study
takes drugs from 3 or more different groups and $y_{ij}=0$ otherwise.  
Writing $p_{ij}$ for the probability that $y_{ij}=1$, the model is 
$$\text{logit}(p_{ij})=x_{ij}^\top\beta+b_i,$$
where we consider normal and $t$ priors for $b_i$, similarly to the six cities
example, with the same hyperprior on the hyperparameter $\zeta$.  
The covariates $x_{ij}$ include \texttt{Gender} ($1=$male, $0=$female), 
\texttt{Race} ($0=$white, $1=$other), some indicators for different
ranges of number of outpatient mental health visits (denoted $\texttt{MHV[j]}$, $j=1,2,3$) and an indicator for inpatient mental health visits
($0$ for none, $1$ otherwise).  

Appendix D in the supplementary materials shows plots of
variational estimates of means, standard deviations and 
skewness for
the random effects compared to an MCMC benchmark for both 
normal and $t$-distributed random effects, as well as the Monte Carlo estimate of the ELBO versus iteration number;  
this appendix also compares the marginal posterior distributions for fixed
effects and variance parameters for the  different methods. 
Similar observations to the previous example can be made here.  
All the SDGM and copula methods are 
superior to the Gaussian approximation in terms of the ELBO and
estimation of the random effect standard deviations, and the two copula
methods are slightly better than SDGM and SDGM-C 
for estimating the skewness of the random effects.  

\subsection{Epilepsy data}

The epilepsy data \citep{thall+v90} considers epileptic seizures for 59 individuals.  The response is a count of the number of seizures experienced, and the value
for the $i$th individual in the $j$th measurement interval is denoted $y_{ij}$, $i=1,\dots, n$, $j=1,\dots 4$.  Each count is for a two-week period.  There is also a baseline covariate (\texttt{Base}) for all individuals which is the log of 1/4 of the 
number of seizures experienced
for 8 weeks prior to treatment.  It is of interest to compare the 
seizure rate between a treatment group given the drug Progabide (\texttt{Trt}=1)
versus a control group (\texttt{Trt}=0).  The response is modelled as Poisson, with mean $\mu_{ij}$, such that 
$$\log \mu_{ij}=x_{ij}^\top \beta+z_{ij}^\top b_i,$$
where $x_{ij}$ are covariates with fixed effects $\beta$ and $z_{ij}$ are
covariates with random effect $b_i$ for subject $i$.  
The prior for $\beta$ is $\beta\sim N(0,100 I)$.  The covariates
$x_{ij}$ include \texttt{Base}, \texttt{Trt}, \texttt{Visit} (coded as $-0.3$ for $j=1$, $-0.1$ for $j=2$, $0.1$ for $j=3$ and $0.3$ for $j=4$), and $\texttt{Base}\times \texttt{Trt}$.  For the random effects, $z_{ij}$ includes an intercept and $\texttt{Visit}$.  Two priors are considered for the random effects $b_i$. The first 
is normal $N(0,\Sigma)$ and the second is $t_{10}(0,\Sigma)$;  in both
cases we write $\Sigma=BB^\top$ where $B$ is the 
Cholesky factor of $\Sigma$ and use a normal $N(0,100 I)$ prior for the elements of $\text{vech}(B)$ after transforming diagonal elements to the log scale.

\begin{figure}[htp]
\begin{center}
\begin{tabular}{c}
\includegraphics[height=90mm]{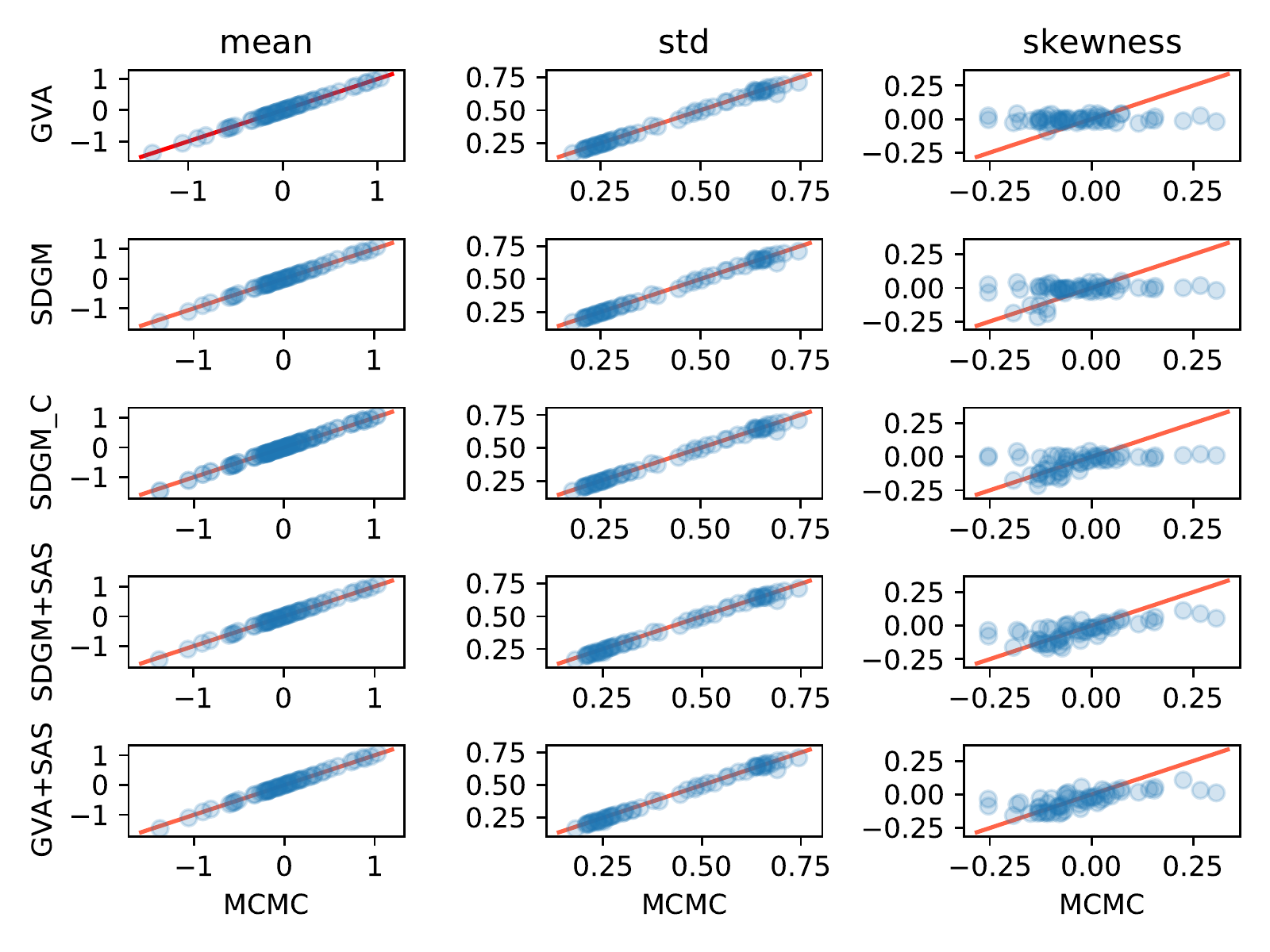} \\
\includegraphics[height=90mm]{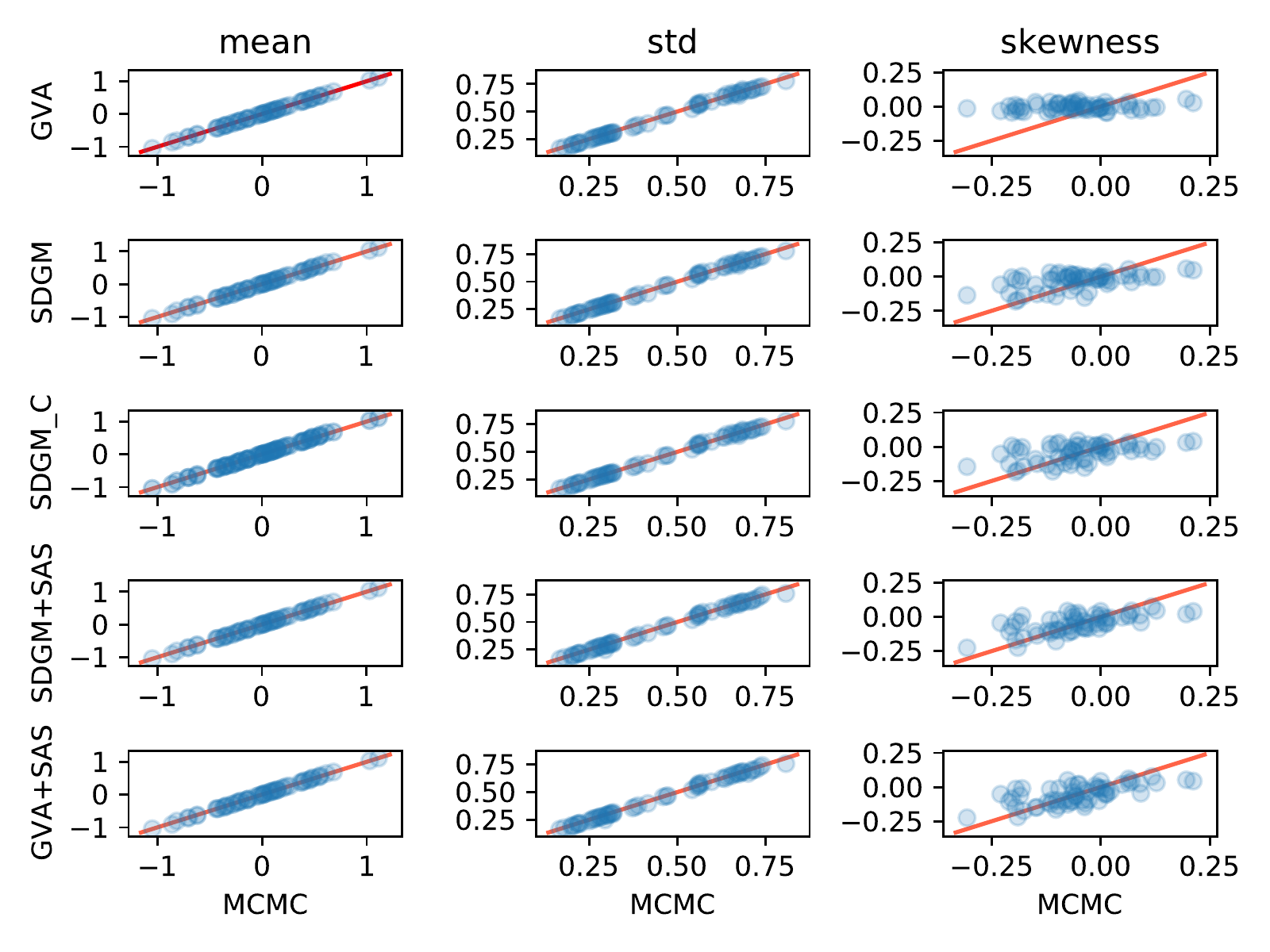}
\end{tabular}
\end{center}
\caption{\label{epilepsy-normal} 
Comparison of mean, standard deviation and skewness estimated by MCMC and approximate methods for random intercept (top) and random slope (bottom) for epilepsy data and normal random effects.  }
\end{figure}

Figure \ref{epilepsy-normal} compares estimates of the means, standard deviations and skewness of the random effects for variational methods versus an MCMC benchmark and normal random effects.  
The figures in Appendix D show a similar plot for the case of $t$-distributed random effects, 
 a plot of the Monte Carlo estimate of the 
ELBO versus iteration number for the normal and $t$-distributed random effects models, and plots of the marginal posterior density estimates
of the fixed effects and variance parameters.

The figures show that the copula methods  capture the skewness of the random effects more accurately than the other methods.  All the methods have 
similar  ELBO values except for the Gaussian approximation, which is the worst. 
Despite the superior performance of the copula methods for estimating
 skewness, this is not reflected in the ELBO.

\subsection{New York stock exchange data}

This example considers variational inference for a stochastic volatility model expressed in state space form.   The example
is also considered in \cite{tan+bn20}, and the data $y$
are modelled as 
\begin{align*}
 y_i & = \exp\left(\sigma b_i+\kappa\right)\epsilon_i,
\end{align*}
where $\epsilon_i$ are iid $N(0,1)$, and $\sigma>0$ and $\kappa$
are real-valued parameters.  The states $b_i$ follow a 
stationary AR(1) model,
\begin{align*}
  b_i & = \phi b_{i-1}+ \gamma_i,
\end{align*}
where $\gamma_i$ are iid $N(0,1)$ and $b_1\sim N(0,1/(1-\phi^2))$.  We  follow \cite{tan+bn20} to obtain an  unconstrained
parametrization by using the  transformations
$$\alpha=\log(\exp(\sigma)-1),\;\;\; \psi=\log \frac{\phi}{1-\phi},$$
so that the model has global parameters $\eta=(\alpha,\psi,\kappa)$, with the states as the local
variables, $b=(b_1,\dots, b_n)$, and $\theta=(b^\top,\eta^\top)^\top$.
Similarly to \cite{tan+bn20}, we use independent $N(0,10)$ priors
for $\alpha$, $\kappa$ and $\psi$.  
The real data used is the New York Stock Exchange (NYSE) data 
available in the R package \texttt{astsa} \citep{stoffer+p23}.  
The data are 100 times mean centred returns over the period 
February 2, 1984 to December 31, 1991.

\begin{figure}[htp]
\begin{center}
\begin{tabular}{c}
\includegraphics[height=90mm]{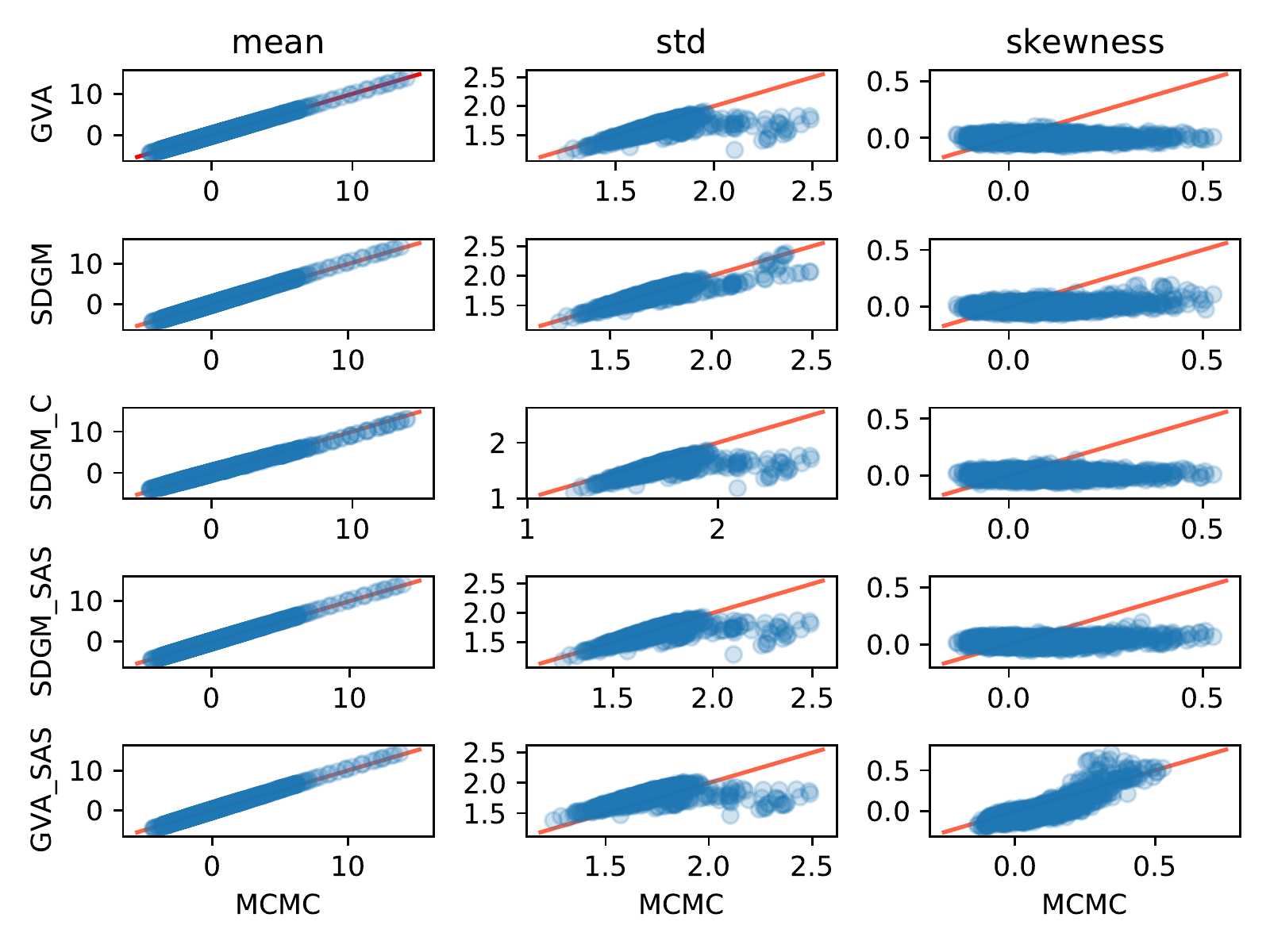} \\
\includegraphics[height=90mm]{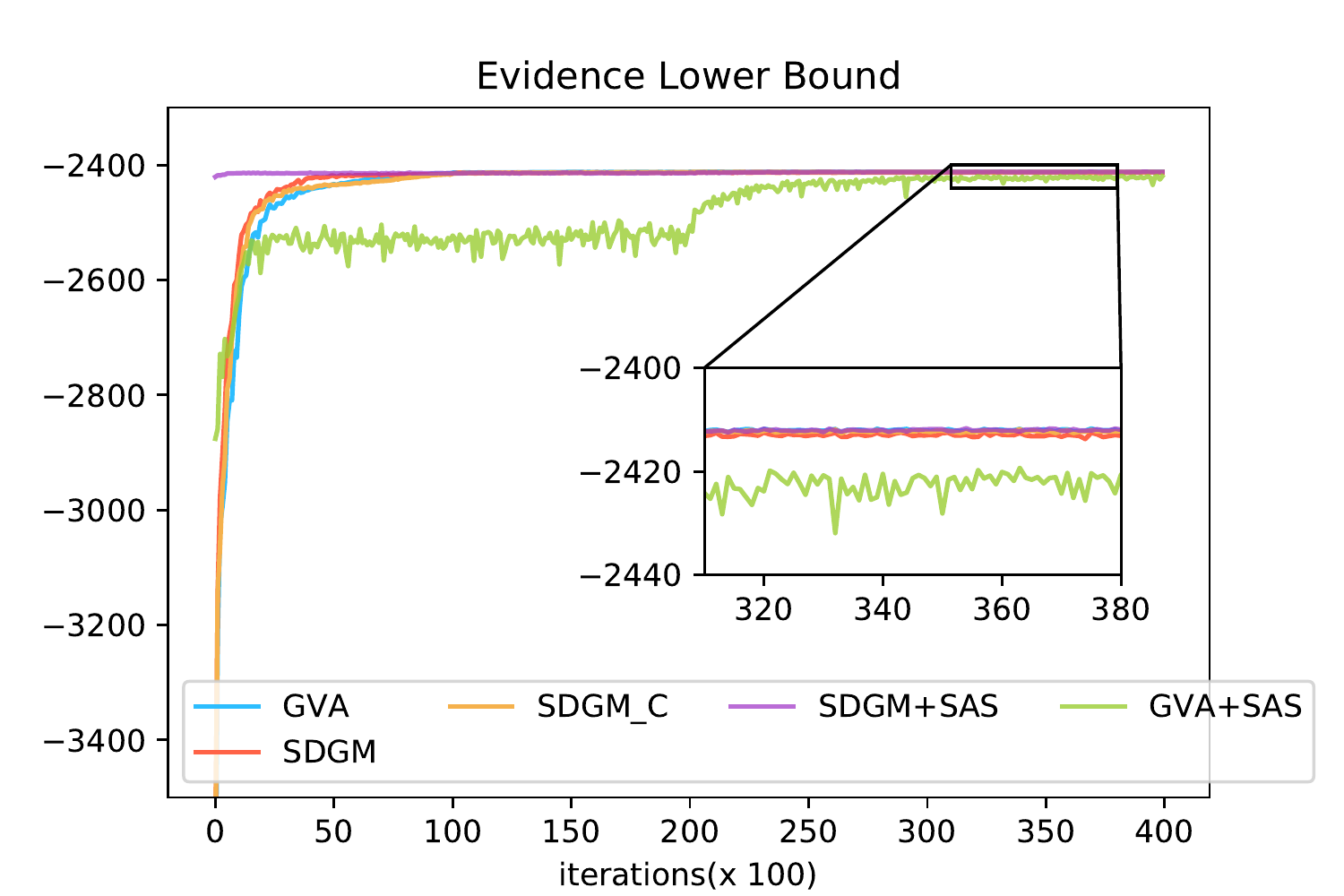} \\
\end{tabular}
\end{center}
\caption{\label{NYSE} 
Comparison of mean, standard deviation and skewness of states estimated by MCMC and approximate methods (top) and Monte Carlo estimate of ELBO versus iteration number (bottom) for NYSE data.  }
\end{figure}

Figure \ref{NYSE} (top) shows the estimation quality
of the marginal posterior means, standard deviations, and
skewness for the states for the various methods.  
Only the GVA-SAS method is able to capture the marginal
skewness, and this could only be achieved with careful
initialization of the optimization.  We tried several different initializations, and ended up doing the following.  In the copula methods, we first fix $\mu$, $L$ and $\kappa$ 
with the GVA results, and then optimize $\alpha$ and the copula parameters $\delta$ and $\epsilon$ for the first 20000 iterations;  following this, we then fix $\alpha$ and the copula parameters, and optimize $\mu$, $L$ and $\kappa$ for another 20000 iterations. This strategy helps GVA+SAS to improve state estimation, but Figure \ref{NYSE} (bottom) shows
the ELBO plot which indicates that even if state estimation is improved for the GVA-SAS method, the achieved lower bound  is
slightly worse.  The posterior marginal densities
for $\alpha$, $\kappa$ and $\psi$ in Appendix D also
demonstrate that GVA-SAS performs poorly for estimating
the global parameters.  Unlike the random
effects examples, we found it difficult to improve on Gaussian variational inference in terms of the ELBO.

\subsection{Computation time}

Tables 1-4 show computation times for the variational
methods considered compared to MCMC.  For all the variational
methods, computation times are based on 50,000 iterations
and the implementation uses automatic differentiation
with \verb|PyTorch|, except for the methods using
sparse linear algebra indicated in Table 4.  
The MCMC results are based on 50,000
iterations in \verb+rstan+ \citep{stan}.  
All computations are performed on a machine with
Intel i7-11800H CPU with 8 cores.  
The variational methods provide speedups in all cases, 
by roughly a factor of 2-10 over the three examples.  
Among the variational approaches, 
the SDGM and SDGM-C methods have similar computational 
demands to GVA, and are somewhat faster
than the copula methods.  For the GVA, SDGM and 
SDGM-C methods for
the state space example, 
we implemented explicit calculation of gradients
using sparse linear algebra, as described in the 
Appendix.  This results in a roughly three-fold speed up
in computation time compared to
an automatic differentiation (AD) implementation in
\verb|PyTorch|.  However, the AD approach was faster for the
random effects examples (results not shown).

\begin{table}[H]
\caption{Computation time - Six Cities data}
\begin{tabular}{|c|ccccc|c|}
\hline
\multirow{2}{*}{\begin{tabular}[c]{@{}c@{}}Time \\ (seconds)\end{tabular}} & \multicolumn{5}{c|}{Variational Approximations (50000 iter)}                                                            & \multirow{2}{*}{\begin{tabular}[c]{@{}c@{}}MCMC\\ (50000 iter)\end{tabular}} \\ \cline{2-6}
                                                                         & \multicolumn{1}{c|}{GVA} & \multicolumn{1}{c|}{SDGM} & \multicolumn{1}{c|}{SDGM\_C} & \multicolumn{1}{c|}{SDGM+SAS} & GVA+SAS &                                                                                    \\ \hline
\begin{tabular}[c]{@{}c@{}}Normal\\ random effects\end{tabular}           & \multicolumn{1}{c|}{80} & \multicolumn{1}{c|}{99}  & \multicolumn{1}{c|}{103}     & \multicolumn{1}{c|}{145}      & 117     & 204                                                                                \\ \hline
\begin{tabular}[c]{@{}c@{}}t-distributed\\ random effects\end{tabular}    & \multicolumn{1}{c|}{80} & \multicolumn{1}{c|}{99}  & \multicolumn{1}{c|}{105}     & \multicolumn{1}{c|}{142}      & 118     & 191                                                                                \\ \hline
\end{tabular}
\end{table}

\begin{table}[H]
\caption{Computation time - Polypharmacy data}
\begin{tabular}{|c|ccccc|c|}
\hline
\multirow{2}{*}{\begin{tabular}[c]{@{}c@{}}Time\\ (seconds)\end{tabular}} & \multicolumn{5}{c|}{Variational Approximations (50000 iter)}                                                            & \multirow{2}{*}{\begin{tabular}[c]{@{}c@{}}MCMC\\ (50000 iter)\end{tabular}} \\ \cline{2-6}
                                                                         & \multicolumn{1}{c|}{GVA} & \multicolumn{1}{c|}{SDGM} & \multicolumn{1}{c|}{SDGM\_C} & \multicolumn{1}{c|}{SDGM+SAS} & GVA+SAS &                                                                                    \\ \hline
\begin{tabular}[c]{@{}c@{}}Normal\\ random effects\end{tabular}           & \multicolumn{1}{c|}{97} & \multicolumn{1}{c|}{120}  & \multicolumn{1}{c|}{123}     & \multicolumn{1}{c|}{159}      & 134     & 616                                                                                \\ \hline
\begin{tabular}[c]{@{}c@{}}t-distributed\\ random effects\end{tabular}    & \multicolumn{1}{c|}{97} & \multicolumn{1}{c|}{120}  & \multicolumn{1}{c|}{120}     & \multicolumn{1}{c|}{158}      & 134     & 626                                                                                \\ \hline
\end{tabular}
\end{table}

\begin{table}[H]
\caption{Computation time -  Epilepsy data}
\begin{tabular}{|c|ccccc|c|}
\hline
\multirow{2}{*}{\begin{tabular}[c]{@{}c@{}}Time\\ (seconds)\end{tabular}} & \multicolumn{5}{c|}{Variational Approximations (50000 iter)}                                                            & \multirow{2}{*}{\begin{tabular}[c]{@{}c@{}}MCMC\\ (50000 iter)\end{tabular}} \\ \cline{2-6}
                                                                          & \multicolumn{1}{c|}{GVA} & \multicolumn{1}{c|}{SDGM} & \multicolumn{1}{c|}{SDGM\_C} & \multicolumn{1}{c|}{SDGM+SAS} & GVA+SAS &                                                                                    \\ \hline
\begin{tabular}[c]{@{}c@{}}Normal\\ random effects\end{tabular}           & \multicolumn{1}{c|}{42} & \multicolumn{1}{c|}{57}  & \multicolumn{1}{c|}{63}     & \multicolumn{1}{c|}{96}      & 74     & 546                                                                                \\ \hline
\begin{tabular}[c]{@{}c@{}}t-distributed\\ random effects\end{tabular}    & \multicolumn{1}{c|}{45} & \multicolumn{1}{c|}{61}  & \multicolumn{1}{c|}{67}     & \multicolumn{1}{c|}{103}      & 80     & 577                                                                                \\ \hline
\end{tabular}
\end{table}

\begin{table}[H]
  \begin{threeparttable}
\caption{Computation time -  NYSE data}
\begin{tabular}{|c|ccccc|c|}
\hline
\multirow{2}{*}{\begin{tabular}[c]{@{}c@{}}Time\\ (seconds)\end{tabular}} & \multicolumn{5}{c|}{Variational Approximations (40000 iter)}                                                            & \multirow{2}{*}{\begin{tabular}[c]{@{}c@{}}MCMC\\ (50000 iter)\end{tabular}} \\ \cline{2-6}
                                                                          & \multicolumn{1}{c|}{GVA} & \multicolumn{1}{c|}{SDGM} & \multicolumn{1}{c|}{SDGM\_C} & \multicolumn{1}{c|}{SDGM+SAS} & GVA+SAS &                                                                                    \\ \hline
\begin{tabular}[c]{@{}c@{}}Normal\\ states\end{tabular}           & \multicolumn{1}{c|}{2413/$858^*$} & \multicolumn{1}{c|}{2528/$851^*$}  & \multicolumn{1}{c|}{2464/$892^*$}     & \multicolumn{1}{c|}{1562}      & 1443     & 2162                                                                                \\ \hline
\end{tabular}
    \begin{tablenotes}
      \small
      \item Times marked with * are for an implementation using sparse matrix computation.
    \end{tablenotes}
  \end{threeparttable}
\end{table}

\section{Discussion}

A new family of variational approximations is introduced that is suitable when the parameter dimension is high and the posterior has known conditional independence structure.
It is based on skew decomposable graphical models, with
the required conditional independence structure  imposed
through sparsity in the precision matrix, similarly to the Gaussian case.  
We explore an alternative centred parametrization of this family which
facilitates an implicit copula extension based on elementwise
transformation of an approximately standardized SDGM random vector.  
Even in the case of an implicit Gaussian copula, our parametrization 
is novel.  The implicit Gaussian
copula and implicit SDGM copula approximations work best, and generally perform similarly.   However, the SDGM and SDGM-C approximations perform nearly as well as the copula methods, but are less computationally demanding. 
Optimization is easier for the copula methods, with less
sensitivity to the choice of learning rates.  

\section*{Acknowledgements} 

Robert Kohn was partially supported by the Australian Research Council grants DP210103873 and IC190100031. 

\bibliographystyle{apalike}
\addcontentsline{toc}{section}{\refname}
\bibliography{skew_normal}

\newpage

\section*{Appendix A - reparametrization gradients for the SDGM family}

 Automatic differentiation is used via the \verb|PyTorch| package in Python \citep{paszke2017}
 in the experiments reported in the main body of the manuscript. The (transposed) vector-Jacobian products (VJPs) given in  \eqref{mc-grad-est}, \eqref{mc-grad-est2}, and    \eqref{mc-grad-est3} are automatically computed by setting $z = \nabla_\theta \log h(\theta)-\nabla_\theta \log q_\lambda(\theta)$ and performing reverse-mode automatic differentiation to obtain the required variational parameter gradient estimates. The gradients $\nabla \log h(\theta)$ and $\nabla \log q(\theta)$ are also obtained via automatic differentiation.  In our random effects examples,  this approach is typically computationally faster than that of a fully sparse matrix implementation. However, 
for the state space model example of Section 4.4 
sparse matrix methods are
faster by roughly a factor of three, and
we give below the required gradients and VJP expressions for such an implementation. 

We now
establish some suitable notation to express the lower bound gradients below. 
For a vector valued function $f$ with vector valued argument $x$, we write 
\begin{align*}
 \frac{df}{dx} & := \left[\frac{\partial f_i(x)}{\partial x_j}\right],
\end{align*}
where the $i$  and $j$ are the row and column indices respectively, 
for the matrix of partial derivatives of $f$ with respect to the components of $x$.  If $f(\cdot)$ is a scalar, then the above is a row vector, 
so that 
\begin{align*}
\frac{df}{dx} & := \nabla_x f(x)^\top.
\end{align*}
If $f(x)$ or $x$ or both are matrix-valued, then
we define
\begin{align*}
\frac{df}{dx} & := \frac{d \text{vec}(f)}{d \text{vec}(x)}.
\end{align*}

In the variational optimization we transform $\kappa$ to $\bar{\kappa}=\log \kappa$ (with the logarithm applied
elementwise) so that $\kappa=\exp(\bar{\kappa})$,
constraining $\kappa$ to be positive.
Although some elements of $L$ are fixed, we
develop ways of estimating the gradient of a variational lower bound with respect to all the elements of $L$ in what follows, 
as this results in compact analytic expressions where gradients with respect to fixed components are ignored in the optimization
updates.   
Expressions are required for 
\begin{align}
\frac{d\theta}{d\lambda} & = \left[ \frac{d\theta}{d\mu},\frac{d\theta}{d\alpha},\frac{d\theta}{d\kappa},\frac{d\theta}{d L}\right]^\top, \label{dthetadlambda}
\end{align}
and $\nabla_\theta \log q_\lambda(\theta)$ to compute a Monte Carlo estimate of the gradient lower bound using \eqref{repar-grad}.
The expression for $\nabla_\theta \log h(\theta)$ is model specific, and
is derived on a case-by-case basis or computed using automatic differentiation.  
To simplify notation we write 
\begin{align*}
x = x(\kappa, \alpha)  = \alpha \odot \kappa^{-1}\odot ({\bf 1}+\alpha^2)^{-1/2} \odot |U|+\kappa^{-1}\odot ({\bf  1}+\alpha^2)^{-1/2}\odot V, 
\end{align*}
so that a draw $\theta$ from the variational distribution is written as $\theta=\mu+L^{-\top}x$.

For $\nabla_\theta \log q_\lambda(\theta)$, we obtain 
\begin{align} \label{eq:grad_logq}
\nabla_\theta \log q_\lambda(\theta)  = & \nabla_\theta \log \phi(\theta;\mu,\m Q^{-1})  +\sum_{k=1}^p \nabla_\theta \log \Phi\left( \left\{D_\kappa D_\alpha L^\top (\theta-\mu)\right\}_k\right);
\end{align}
where
\begin{align*}
\nabla_\theta \log \phi(\theta;\mu, \m Q^{-1}) & = -\m Q(\theta-\mu) = - L D_{\kappa}^2 L^\top (\theta - \mu).\\
\nabla_\theta \log \Phi\left(\left\{D_\kappa D_\alpha L^\top (\theta-\mu)\right\}_k\right) & = 
 \frac{\phi\left(\left\{D_\kappa D_\alpha L^\top (\theta-\mu)\right\}_k\right)}{\Phi\left(\left\{D_\kappa D_\alpha L^\top (\theta-\mu)\right\}_k\right)}\times \left\{D_\kappa D_\alpha L^\top \right\}_{k.}^\top,
\end{align*}
and $\left\{D_\kappa D_\alpha L^\top \right\}_{k.}$ is the $k$th row of the matrix $D_\kappa D_\alpha L^\top$.  

To obtain an unbiased estimator of \eqref{repar-grad} efficiently based on a single Monte Carlo sample of $(U,V)$, we need
to evaluate, for $\theta=\theta(u,v ; \lambda)$ and $z=\nabla_\theta \log h(\theta)-\nabla_\theta \log q_\lambda(\theta)$, the Jacobian-vector product
\begin{align}
   \frac{d\theta}{d\lambda}^\top z & = \left[ \frac{d\theta}{d\mu}^\top z, \frac{d\theta}{d\alpha}^\top z, 
  \frac{d\theta}{d\bar{\kappa}}^\top z, \frac{d\theta}{d L}^\top z\right].  \label{mc-grad-est}
 \end{align}
By matrix calculus, 
\[
\begin{split}
    \frac{d\theta}{d\mu}^\top z &= z \\ 
    \frac{d\theta}{d\alpha}^\top z &= \bigg[L^{-\top}{\rm diag}\left(\kappa^{-1} \odot ({\bf 1}+\alpha^2)^{-3/2} \odot \left(|U| - \alpha \odot V\right) \right)\bigg]^\top z \\ 
    & =  \kappa^{-1} \odot ({\bf 1}+\alpha^2)^{-3/2} \odot \left(|U| - \alpha \odot V\right) \odot \left(L^{-1} z\right)\\ 
   \frac{d\theta}{d \overline{\kappa}}^\top z &= \ \left[-L^{-\top}{\rm diag}(x)\right]^\top z = -x \odot \left(L^{-1} z\right)  \\
\frac{d \theta}{d L}^\top z &= -\text{vec}(L^{-\top}xz^\top L^{-\top}). \end{split}\]
When $L$ is sparse, these expressions can be evaluated efficiently, because their computation involves sparse triangular linear systems.


\section*{Appendix B - reparametrization gradients for the centered parametrization}

In the variational optimization for the centered parametrization 
we transform $\nu$ to $\bar{\nu}=\log \nu$ (where the logarithm is applied
elementwise) so that $\nu=\exp(\bar{\nu})$,
so that $\nu$ is  positive. 
To compute a Monte Carlo estimate of the gradient lower bound using \eqref{repar-grad} we require expressions for
\begin{align}
\frac{d\theta}{d\rho} & = \left[ \frac{d\theta}{d\xi},\frac{d\theta}{d\alpha},\frac{d\theta}{d\bar{\nu}},\frac{d\theta}{d L}\right]^\top, \label{dthetadlambda2}
\end{align}
and $\nabla_\theta \log q_\rho(\theta)$.  Computing $\nabla_\theta \log h(\theta)$ is model specific. Both computations are done similarly to those for   the direct parametrization.

Writing 
\begin{align*}
x = x(\kappa, \alpha)  = \alpha \odot \kappa^{-1}\odot ({\bf 1}+\alpha^2)^{-1/2} \odot |U|+\kappa^{-1}\odot ({\bf  1}+\alpha^2)^{-1/2}\odot V, 
\end{align*}
 a draw $\theta$ from the variational distribution is $\theta=\mu+L^{-\top}x$.   
 
The expression  $\nabla_\theta \log q_\rho(\theta)$, 
 is computed similarly to corresponding expression in Appendix A after substituting 
$\kappa=\sigma(\alpha)\oslash \nu$, and then $\mu=\xi-L^{-\top}D_\kappa^{-1} \mu(\alpha)$.
To obtain an unbiased estimator of \eqref{repar-grad} efficiently based on a single Monte Carlo sample of $(U,V)$, we need
to evaluate, for $\theta=\theta(u,v ; \rho)$ and $z=\nabla_\theta \log h(\theta)-\nabla_\theta \log q_\rho(\theta)$, the Jacobian-vector product
\begin{align}
   \frac{d\theta}{d\rho}^\top z & = \left[ \frac{d\theta}{d\xi}^\top z, \frac{d\theta}{d\alpha}^\top z, 
  \frac{d\theta}{d\bar{\nu}}^\top z, \frac{d\theta}{d L}^\top z\right].  \label{mc-grad-est2}
 \end{align}
 For the first, third and fourth terms on the right, it is
 straightforward to obtain
\[
\begin{split}
    \frac{d\theta}{d\xi}^\top z &= z \\ 
   \frac{d\theta}{d \overline{\nu}}^\top z &= \text{diag}(\nu\odot Z_\alpha^c) L^{-1} z\\
\frac{d \theta}{d L}^\top z &= -\text{vec}(L^{-\top}(\nu\odot Z_\alpha^c) z^\top L^{-\top}). \end{split}\]

Finally,
\begin{align*} 
      \frac{d\theta}{d\alpha}^\top z &= \frac{d Z_\alpha^c}{d\alpha}^\top \text{diag}(\nu) L^{-1} z,
\end{align*}
where, writing $\text{dg}(A)$ for the vector of diagonal elements of a square matrix
$A$, 
\begin{align*}
  \frac{dZ_\alpha^c}{d\alpha} & = \text{diag}\left\{\left(\sigma(\alpha)\odot \text{dg}\left(\frac{dZ_\alpha^c}{d\alpha}-\frac{d\mu(\alpha)}{d\alpha}\right)-(Z_\alpha-\mu(\alpha))\odot \text{dg}\left(\frac{d\sigma(\alpha)}{ d\alpha}\right)\right)\oslash \sigma(\alpha)^2 \right\};
\end{align*}
\begin{align*}
  \frac{dZ_\alpha}{d\alpha} & = \text{diag}\left(\text{dg}\left(\frac{d\delta(\alpha)}{d\alpha}\right)\odot |U|-\delta \odot V\right),
\end{align*}
\begin{align*}
 \frac{d\mu(\alpha)}{d\alpha} & = \frac{\sqrt{2}}{\pi}\frac{d\delta(\alpha)}{d \alpha},
\end{align*}
\begin{align*}
  \frac{d\sigma(\alpha)}{d\alpha} & = \text{diag}\left(-\frac{2\delta(\alpha)}{\pi}\odot \left(1-\frac{2\delta(\alpha)^2}{\pi}\right)^{-1/2}\odot \text{dg}\left(\frac{d\delta(\alpha)}{d \alpha}\right)\right),
\end{align*}
where
\begin{align*}
  \frac{d\delta(\alpha)}{d\alpha}  & = \text{diag}\left((1+\alpha^2)^{-3/2}\right).
\end{align*}

\section*{Appendix C - reparametrization gradients for SDGM with SAS transformation}

We need $\nabla_{\breve{\lambda}} \log q_{\breve{\lambda}}(\theta)$ to obtain the reparametrization gradients.  Write 
$$\widetilde{\theta}'=\left(\frac{d\widetilde{\theta}_1}{d\theta_1},\dots, \frac{d\widetilde{\theta}_p}{d\theta_p}\right) \;\;\;\;
\mbox{ and }\;\;\;\;\widetilde{\theta}''=\left(\frac{d^2\widetilde{\theta}_1}{d\theta_1^2},\dots, 
\frac{d^2\widetilde{\theta}_p}{d\theta_p^2}\right),$$
where \eqref{dthetajtildedthetaj} gives $d\widetilde{\theta}_j/d\theta_j$ 
and 
$$\frac{d^2\widetilde{\theta}_j}{d\theta_j^2} = h_{\gamma_j}''\left(\frac{\theta_j-\xi_j}{\exp(\bar{\nu}_j)}\right)\times \frac{1}{\exp(2\bar{\nu}_j)}.$$
Using \eqref{qsas}, 
\begin{align*}
  \nabla_{\theta} \log q_{\breve{\lambda}}(\theta) & = \widetilde{\theta}'\odot \nabla_{\widetilde{\theta}}\log q_\lambda(\widetilde{\theta})+\widetilde{\theta}''\oslash \widetilde{\theta}',
\end{align*}
where $\nabla_{\widetilde{\theta}} \log q_\lambda(\widetilde{\theta})$
is previously
computed (as the gradient of the log of an SDGM density).

For the reparametrization gradients it is also necessary  to compute
Jacobian vector products of the form
\begin{align}
   \frac{d\theta}{d\breve{\lambda}}^\top z & = \left[ \frac{d\theta}{d\xi}^\top z, \frac{d\theta}{d\alpha}^\top z, 
  \frac{d\theta}{d\bar{\nu}}^\top z, \frac{d\theta}{d L}^\top z, \frac{d\theta}{d \gamma}^\top z \right].  \label{mc-grad-est3}
 \end{align}
 Write
 $$t_\gamma'(z)=\left[\frac{dt_{\gamma_1}(z)}{dz},\dots, \frac{dt_{\gamma_p}(z_p)}{dz_p}\right],$$
 and
 $$\dot{t}_\gamma(z)=\left[\frac{dt_{\gamma_1}(z)}{d\gamma_1},\dots, \frac{dt_{\gamma_p}(z_p)}{d\gamma_p}\right].$$
For the terms on the right of \eqref{mc-grad-est3}: 
\[
\begin{split}
    \frac{d\theta}{d\xi}^\top z &= z \\ 
   \frac{d\theta}{d \overline{\nu}}^\top z &= t_\gamma'(L^{-\top}Z_\alpha^c)\odot \exp(\bar{\nu})\odot z \\
   \frac{d\theta}{d\alpha}^\top z & = \frac{dZ_\alpha^c}{d\alpha}^\top 
     L^{-1}(t_\gamma'(L^{-\top}Z_\alpha^c)\odot \exp(\bar{\nu})\odot z) \\
\frac{d \theta}{d L}^\top z &= -\text{vec}\left(L^{-\top}Z_\alpha^c \left\{t_\gamma'(L^{\top}Z_\alpha^c)\odot \exp(\bar{\nu})\odot z\right\}^\top L^{-\top}\right) \\
\frac{d\theta}{d\gamma}^\top z & = \exp(\bar{\nu})\odot \dot{t}_\gamma(L^{-1}Z_\alpha^c)\odot z
\end{split}\]
$\nicefrac{dZ_\alpha^c}{d\alpha}$ is computed
in Appendix B.  The expressions above can be efficiently computed by making
use of the sparsity of $L$.   

\section*{Appendix D - additional figures for examples}

\begin{figure}[htp]
\begin{center}
\begin{tabular}{c}
\includegraphics[height=90mm]{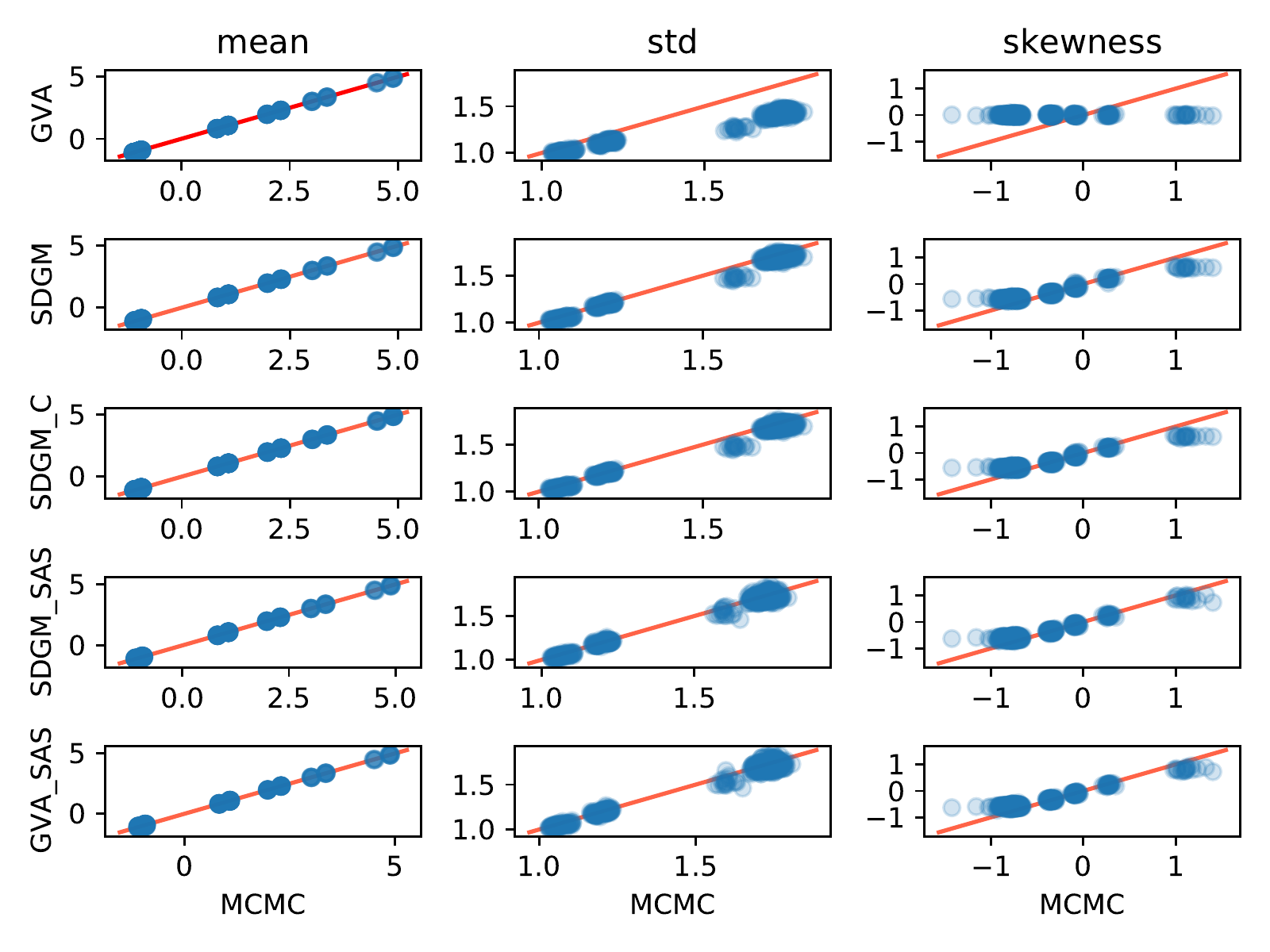}  \\
\includegraphics[height=90mm]{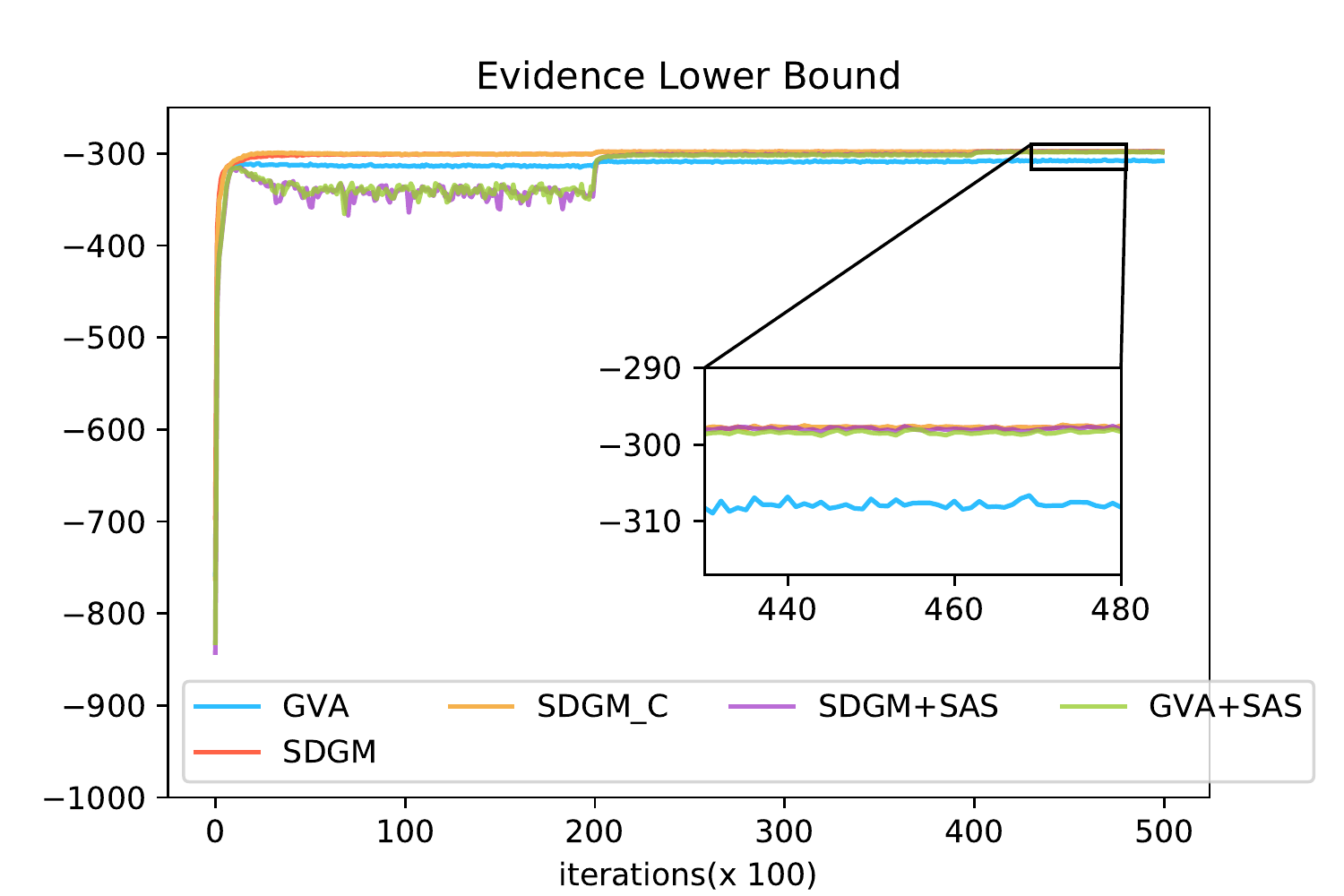} 
\end{tabular}
\end{center}
\caption{\label{six-cities-t} 
Comparison of means, standard deviations and skewness for random effects estimated by MCMC and approximate methods (top) and Monte Carlo estimate of ELBO versus iteration number (bottom) for six cities data and $t$-distributed random effects.  }
\end{figure}

\begin{figure}[H]
\begin{center}
\begin{tabular}{c}
\includegraphics[height=90mm]{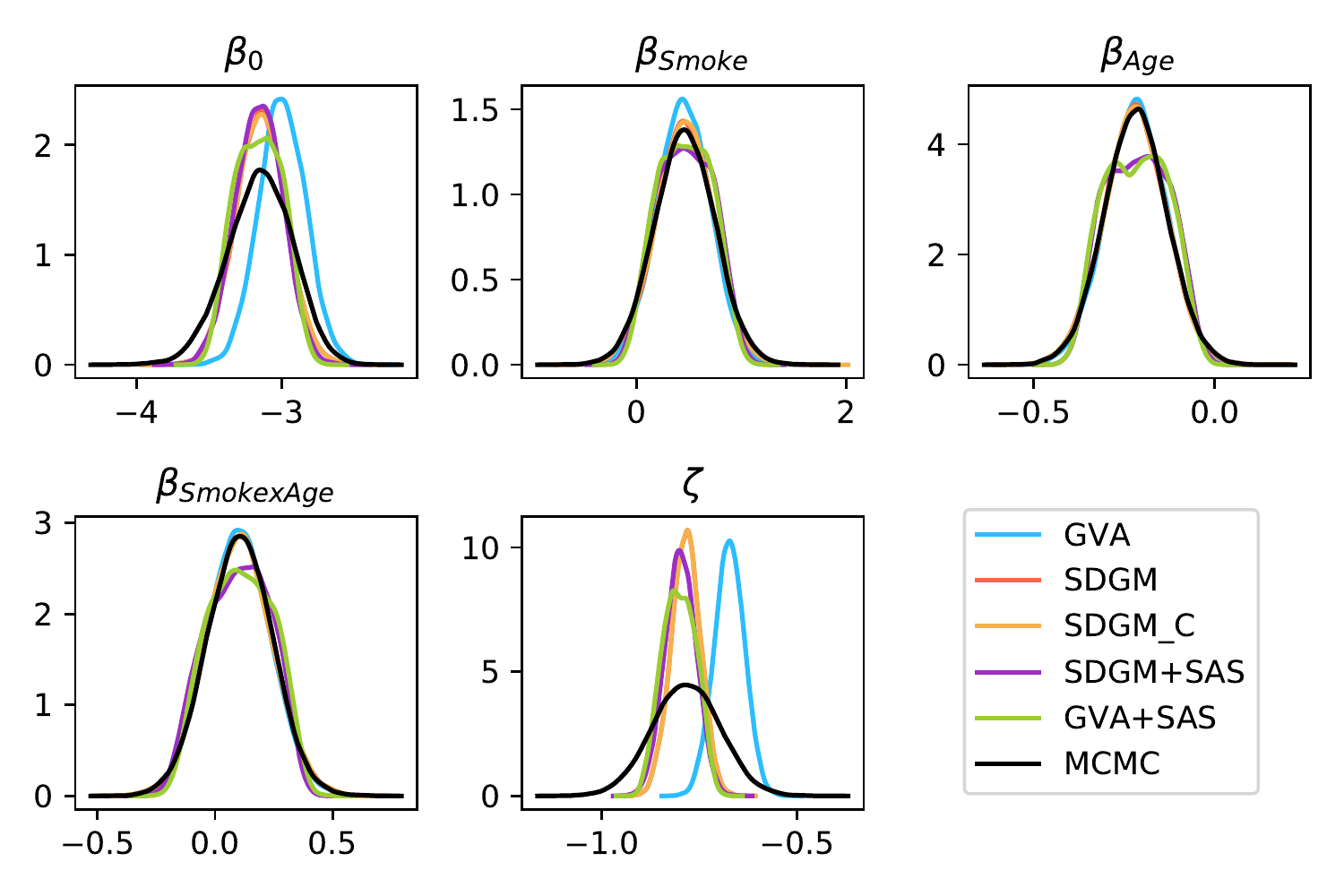} \\
\includegraphics[height=90mm]{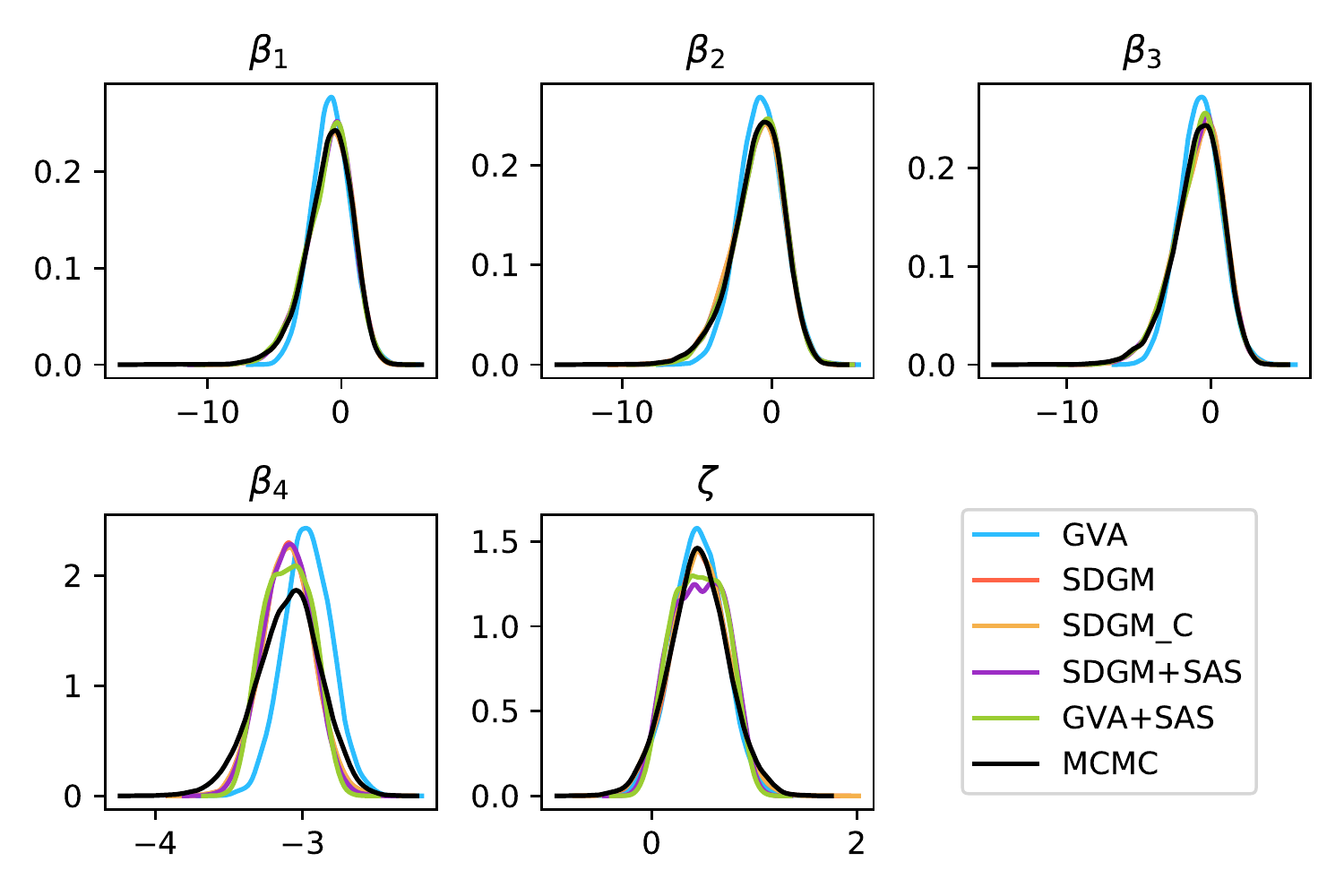}
\end{tabular}
\end{center}
\caption{\label{six-cities-m}Comparison of marginal posterior densities
for fixed effects parameters and variance parameter for six cities data for
normal random effects (top) and $t$-distributed random effects (bottom).}
\end{figure}

Figure \ref{six-cities-t} shows the
estimation quality of the $t$-distributed random effects and the ELBO
plot for the six cities example.   Figure \ref{six-cities-m} compares the performance of the variational methods
versus MCMC for estimating marginal posterior distributions of fixed
effects parameters and variance parameters for the six cities example.  

\begin{figure}[htp]
\begin{center}
\begin{tabular}{c}
\includegraphics[height=90mm]{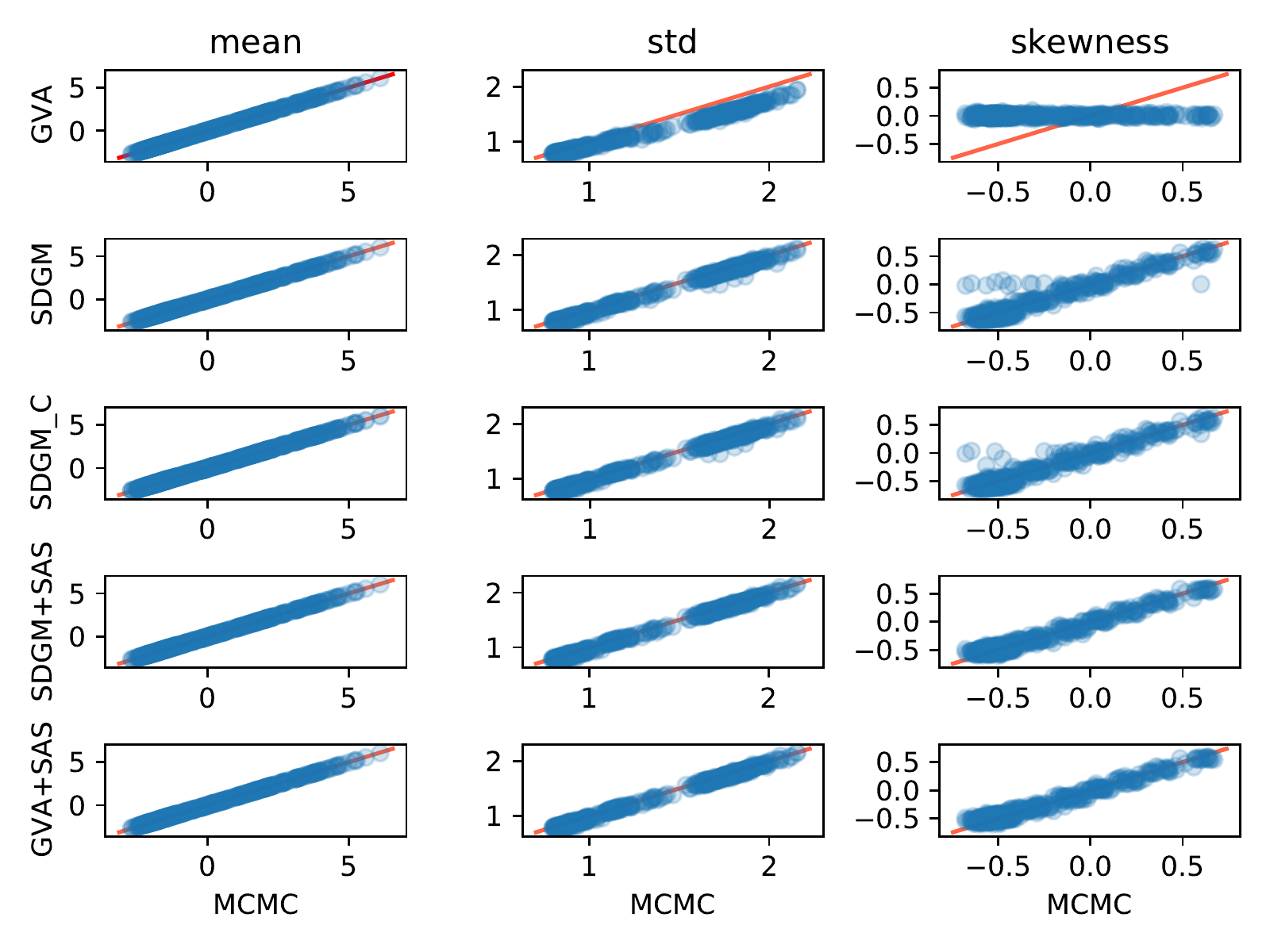} \\
\includegraphics[height=90mm]{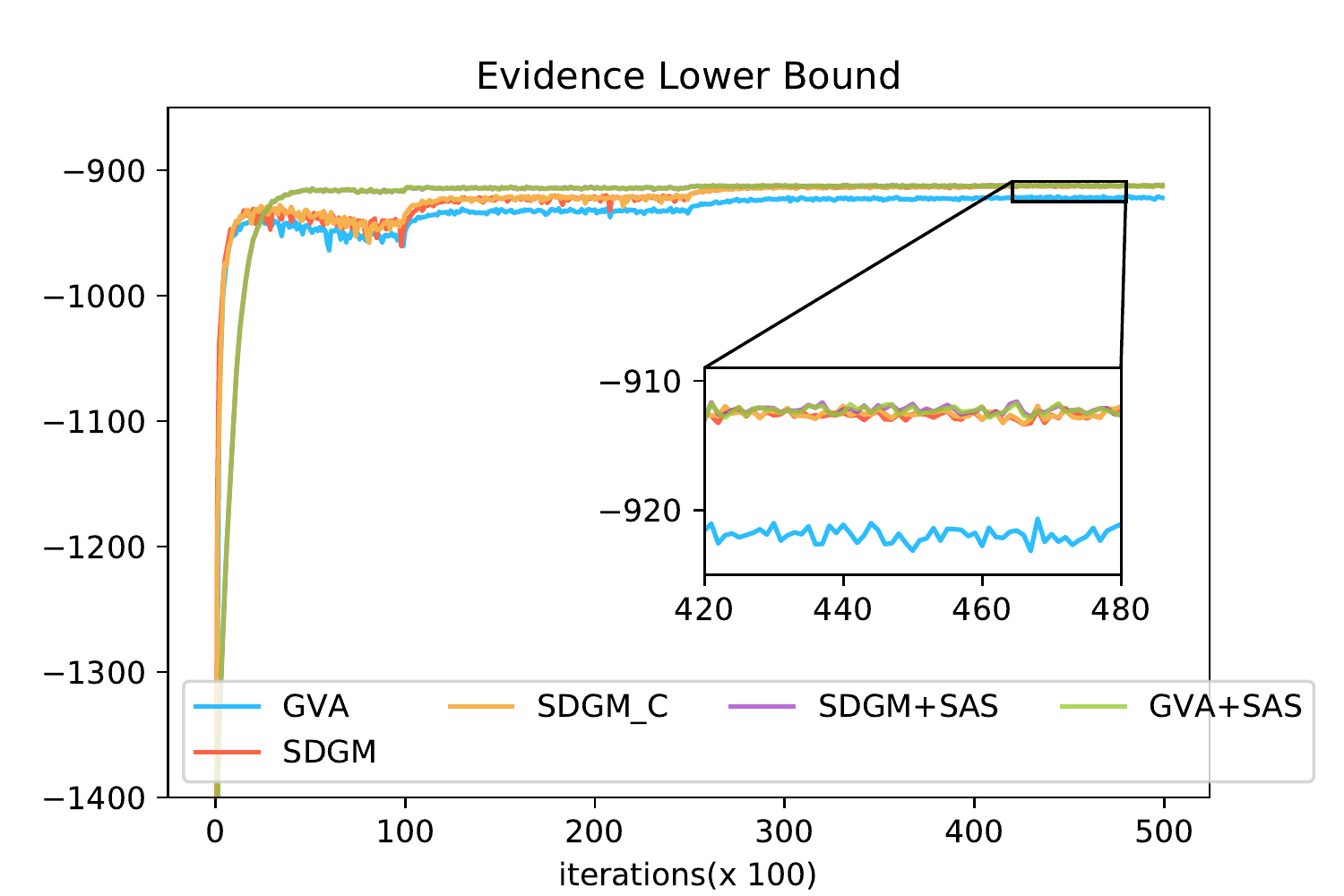} \\
\end{tabular}
\end{center}
\caption{\label{polypharmacy-normal} 
Comparison of mean, standard deviation and skewness estimated by MCMC and approximate methods (top) and Monte Carlo estimate of ELBO versus iteration number (bottom) for polypharmacy data and normal random effects.  }
\end{figure}

\begin{figure}[htp]
\begin{center}
\begin{tabular}{c}
\includegraphics[height=90mm]{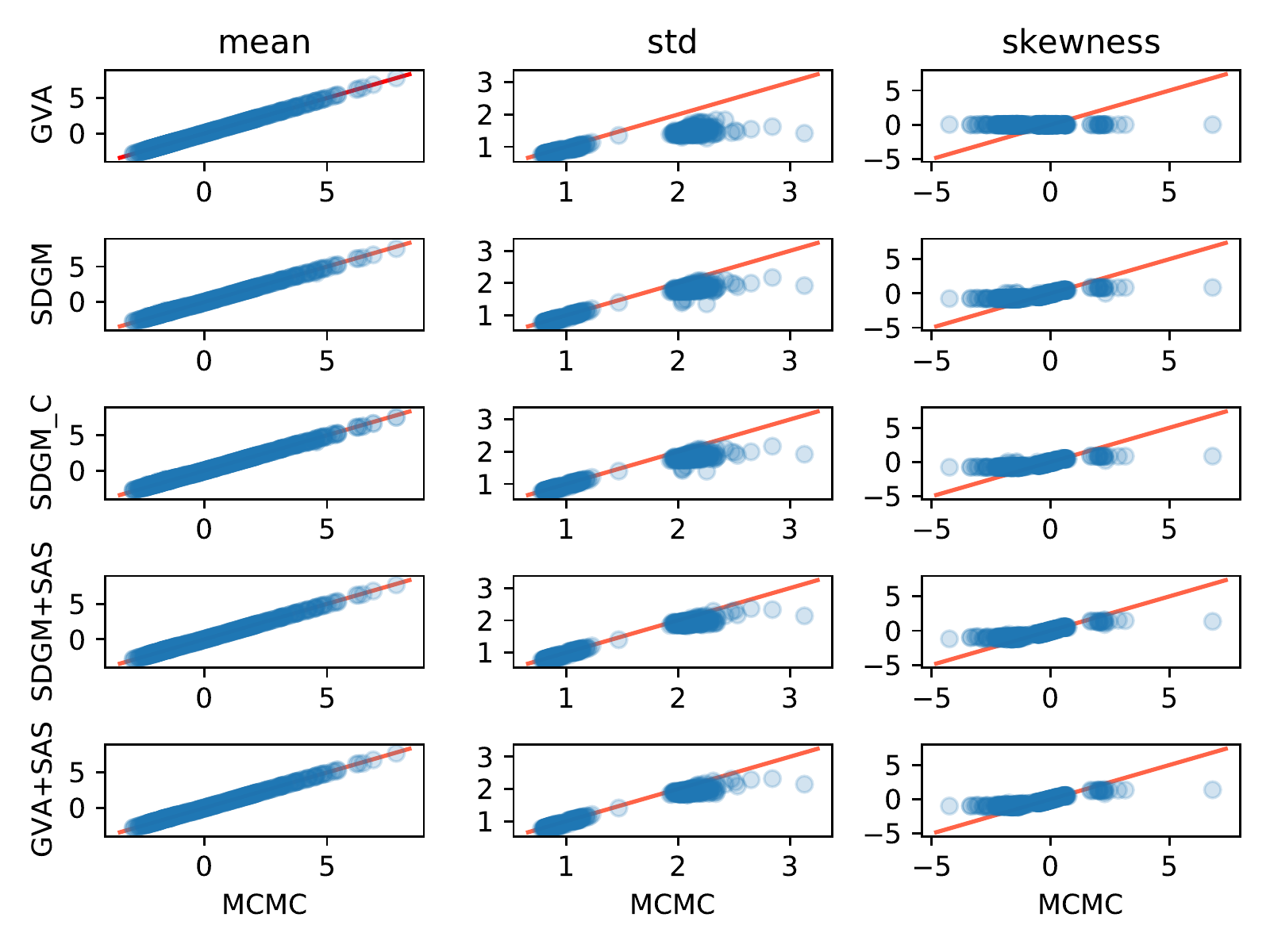}  \\
\includegraphics[height=90mm]{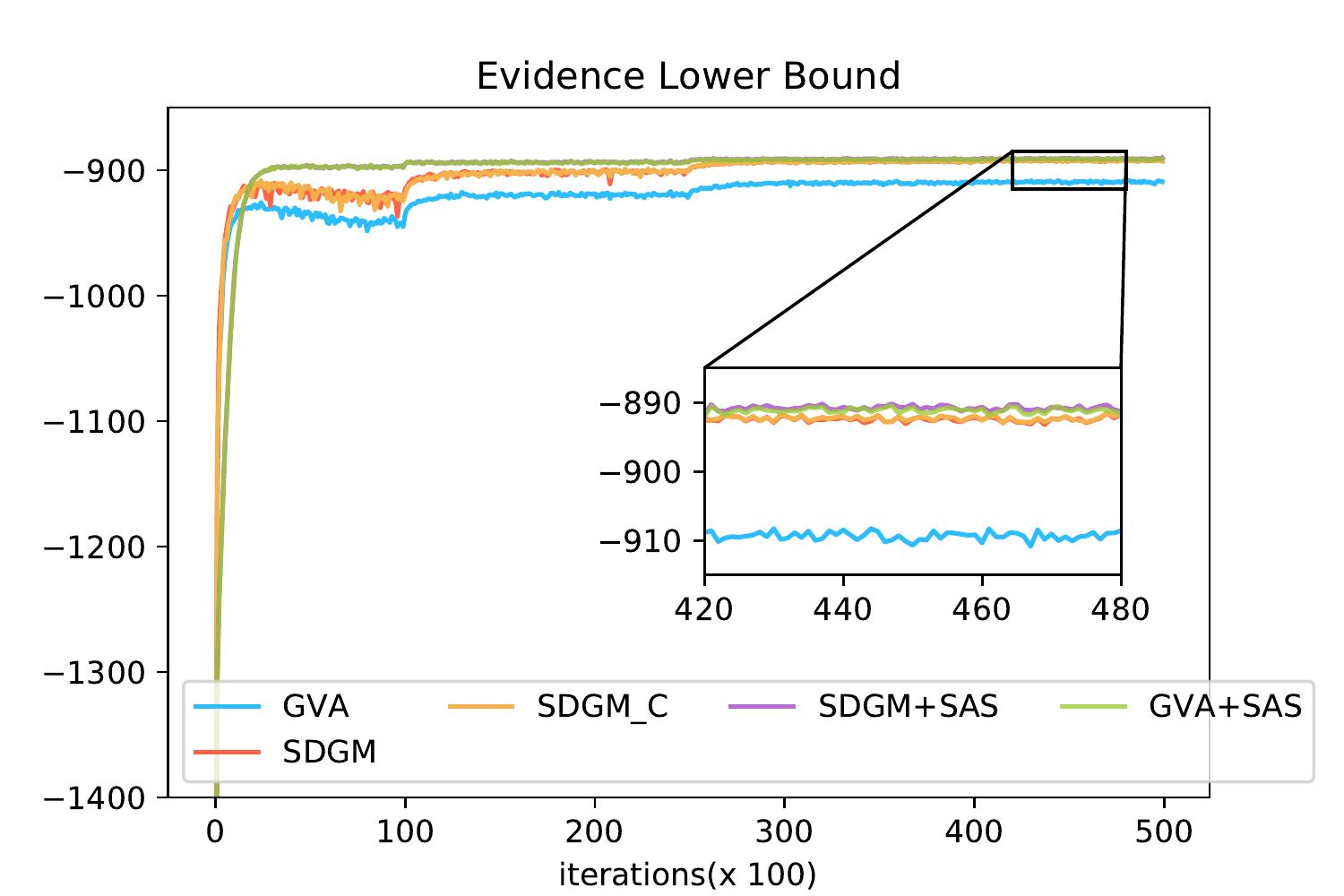} 
\end{tabular}
\end{center}
\caption{\label{polypharmacy-t} 
Comparison of mean, standard deviation and skewness estimated by MCMC and approximate methods (top) and Monte Carlo estimate of ELBO versus iteration number (bottom) for polypharmacy data and $t$-distributed random effects.  }
\end{figure}

\begin{figure}[H]
\begin{center}
\begin{tabular}{c}
\includegraphics[height=90mm]{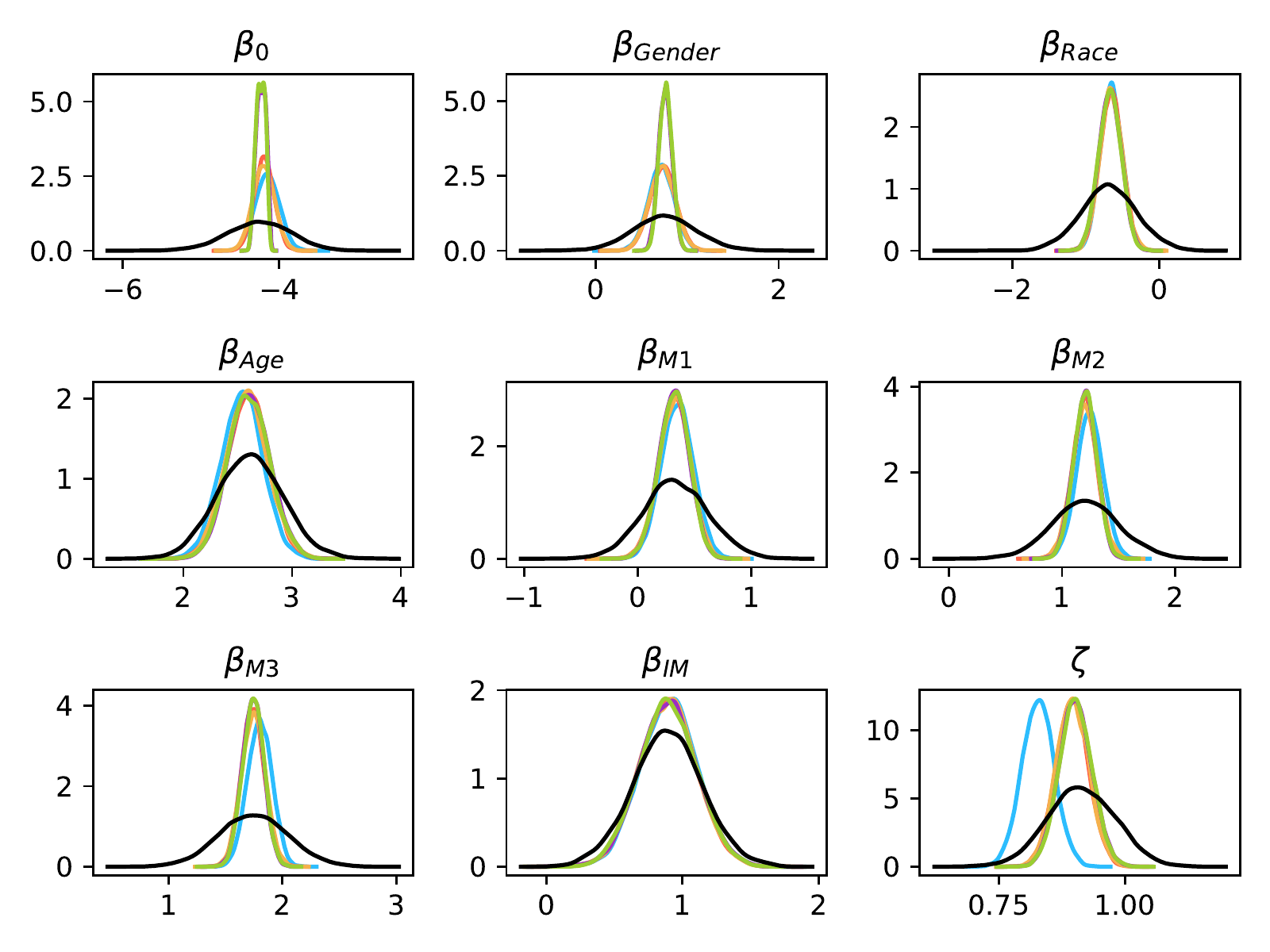} \\
\includegraphics[height=90mm]{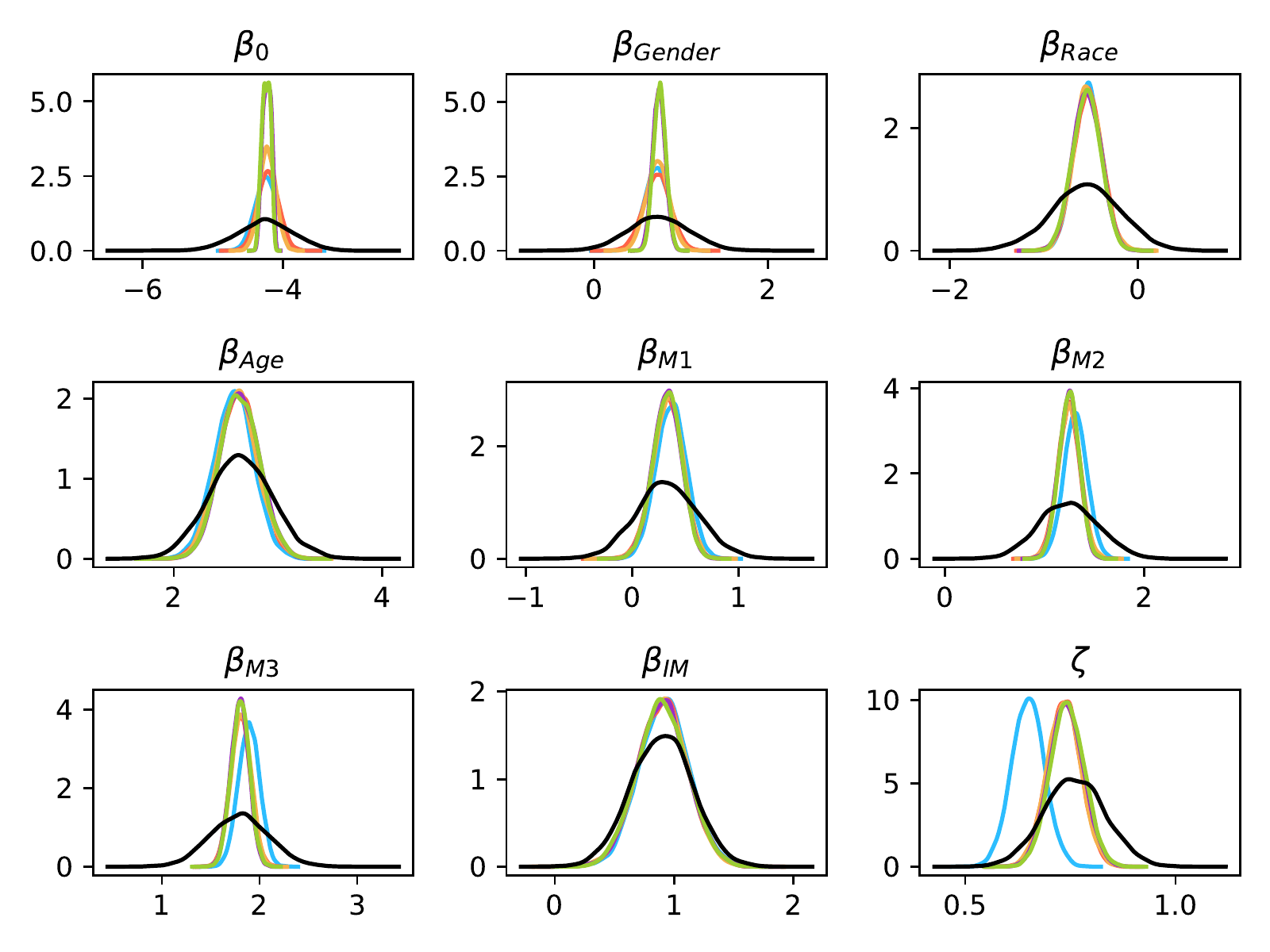}
\end{tabular}
\end{center}
\caption{\label{polypharmacy-m}Comparison of marginal posterior densities
for fixed effects parameters and variance parameter for polypharmacy data for
normal random effects (top) and $t$-distributed random effects (bottom).}
\end{figure}

Figures \ref{polypharmacy-normal} and \ref{polypharmacy-t} shows shows the
estimation quality of the random effects in terms of mean, standard deviation and skewness for the polypharmacy example, and the ELBO versus iteration number, for the cases of normally distributed and $t$-distributed random effects. 
 Figure \ref{polypharmacy-m}, shows the marginal posterior distributions
 for the fixed effects and variance parameter.

\begin{figure}[htp]
\begin{center}
\begin{tabular}{c}
\includegraphics[height=90mm]{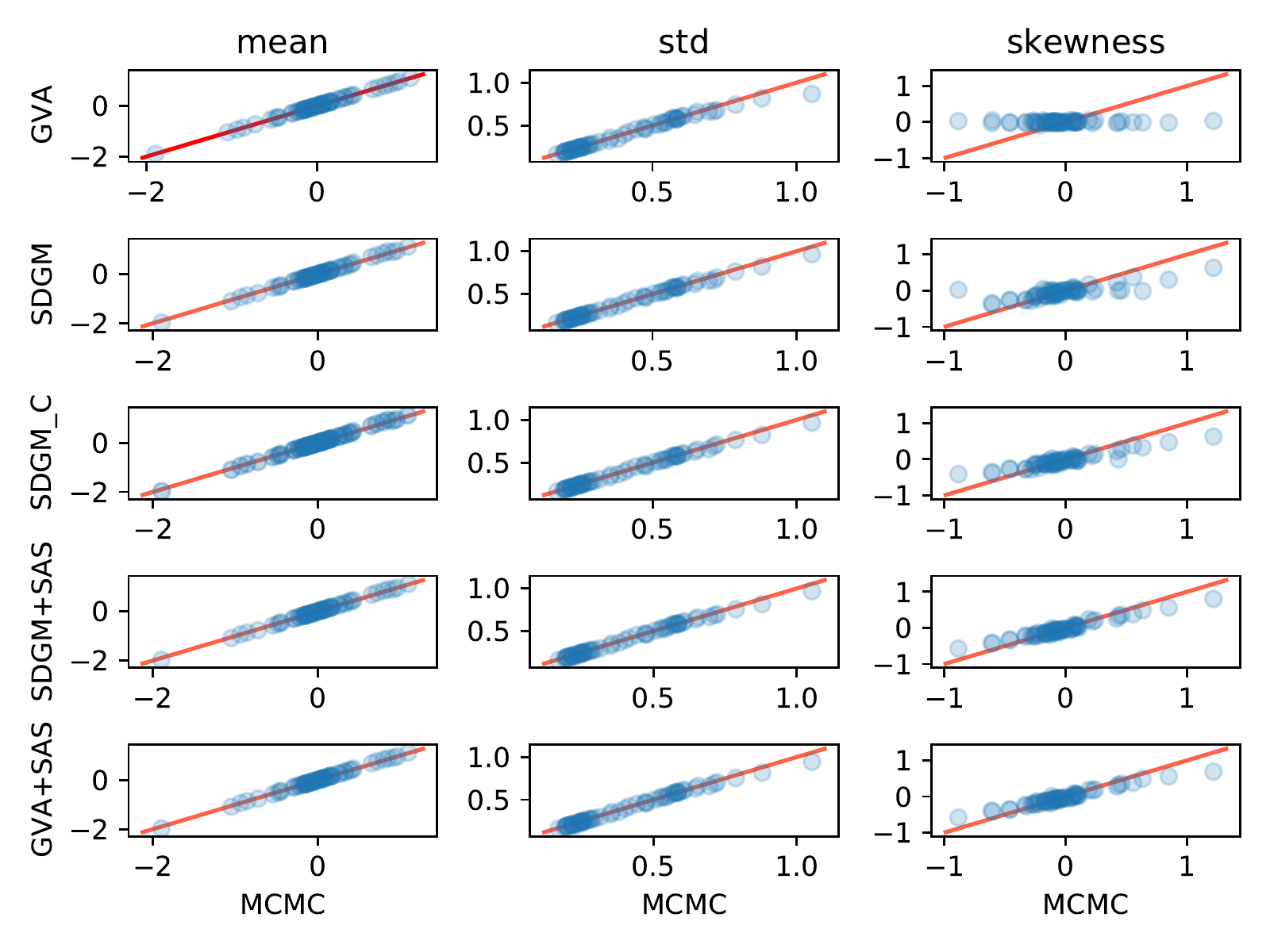} \\
\includegraphics[height=90mm]{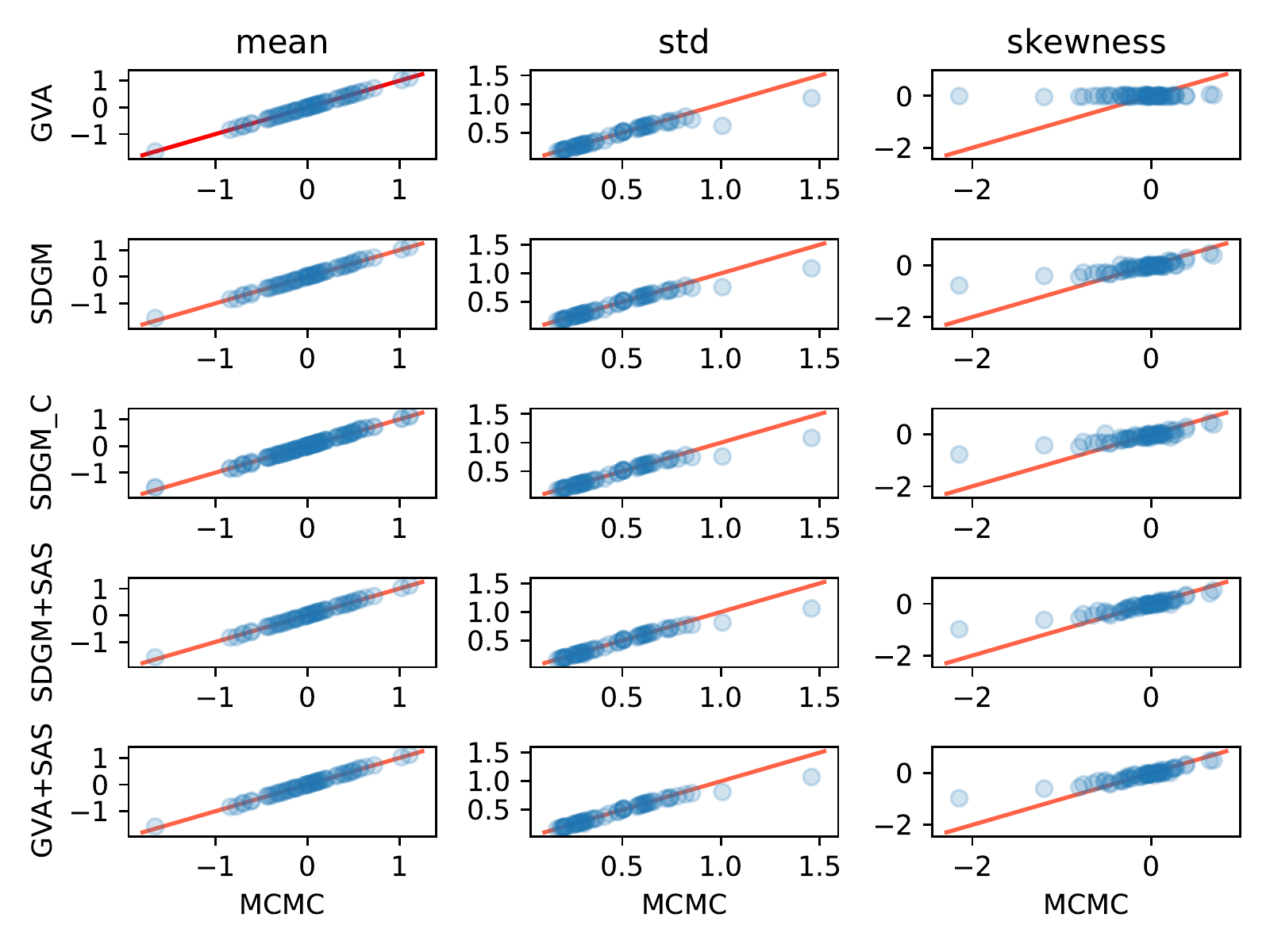} 
\end{tabular}
\end{center}
\caption{\label{epilepsy-t} 
Estimated posterior marginal densities for fixed effects and variance parameter (top) and comparison of mean, standard deviation and skewness estimated by MCMC and approximate methods (bottom) for epilepsy data and $t$-distributed random effects.  }
\end{figure}

\begin{figure}[H]
\begin{center}
\begin{tabular}{c}
\includegraphics[height=90mm]{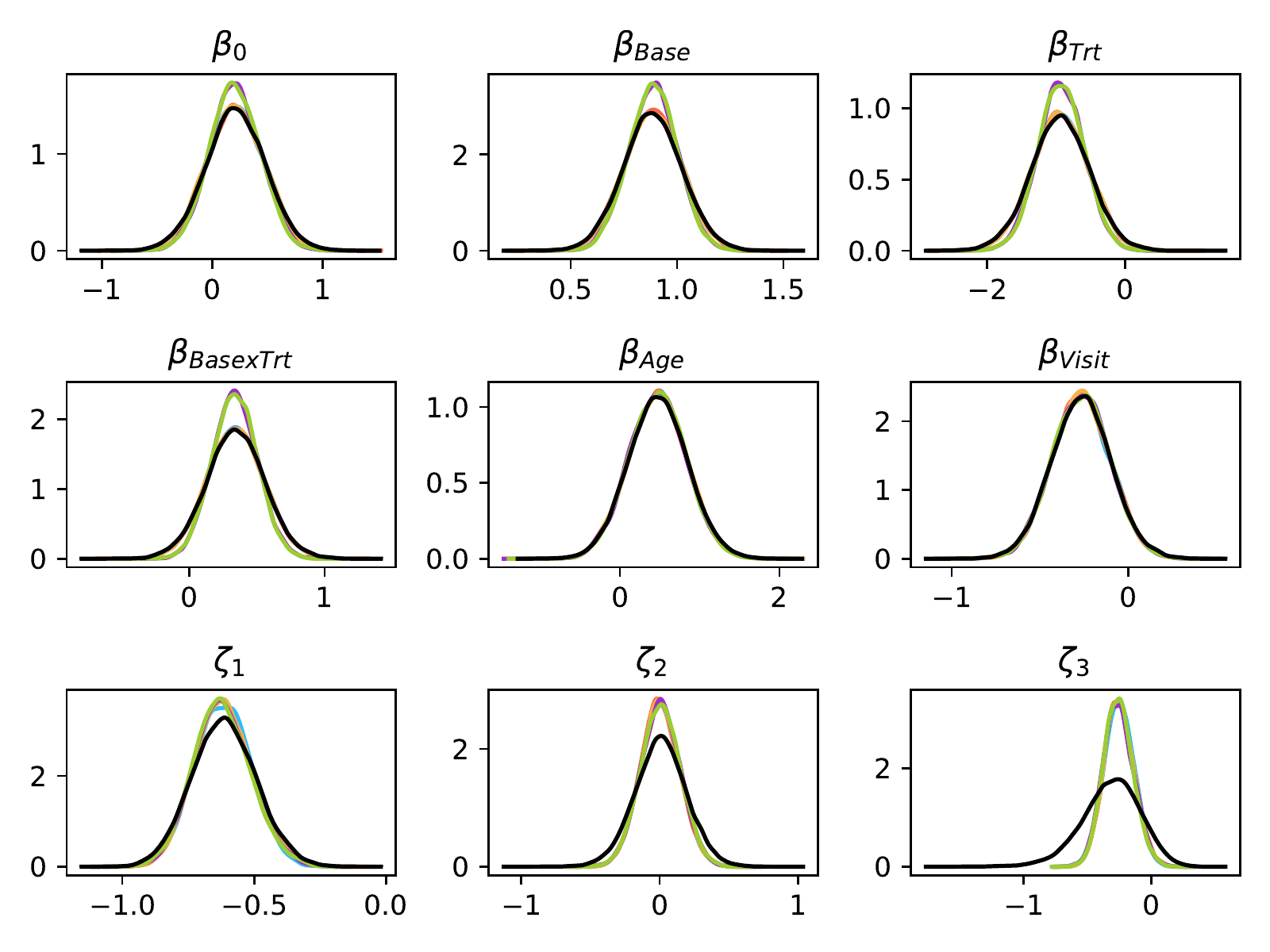} \\
\includegraphics[height=90mm]{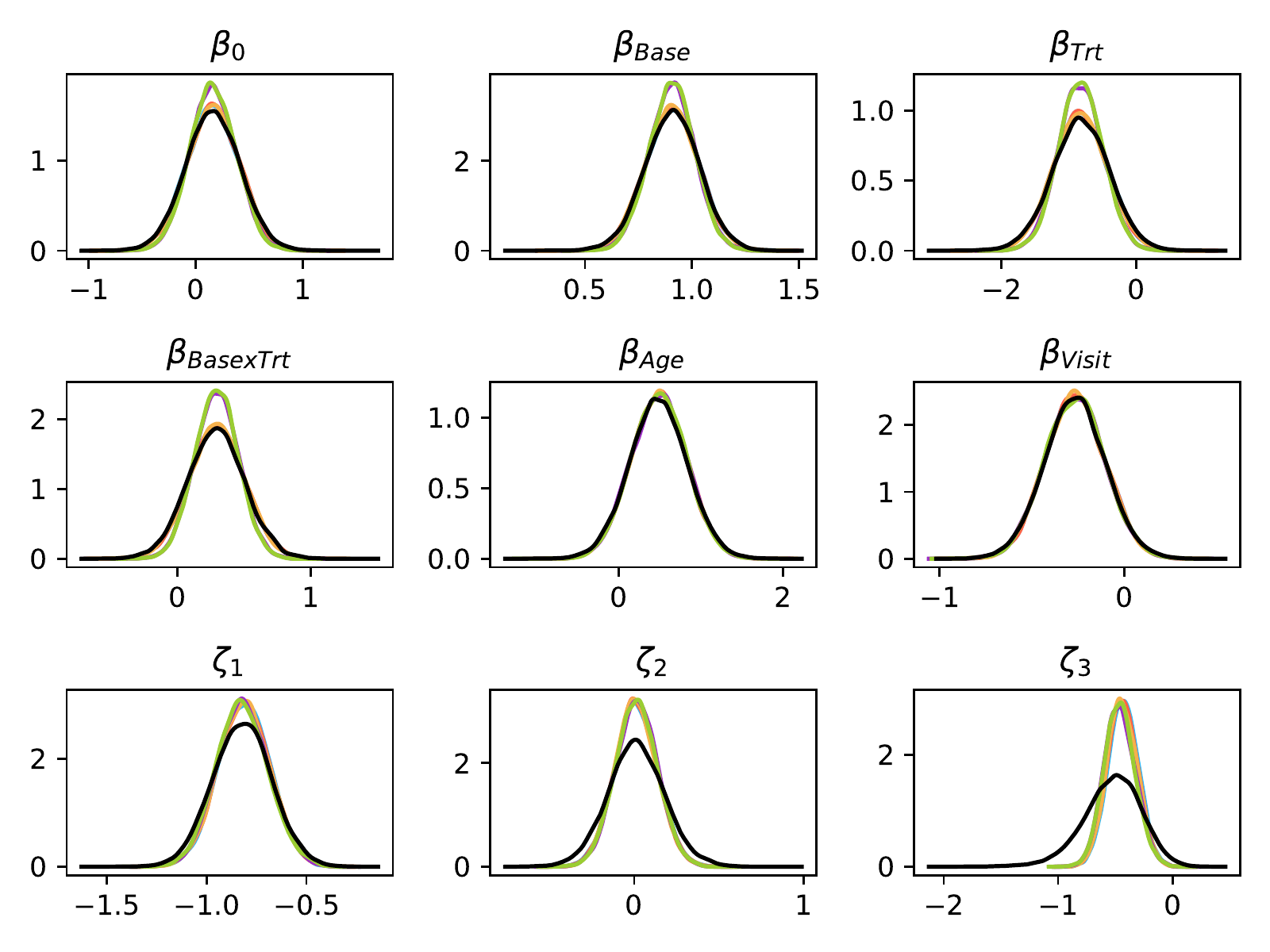}
\end{tabular}
\end{center}
\caption{\label{epilepsy-m}Comparison of marginal posterior densities
for fixed effects parameters and variance parameter for epilelpsy data for
normal random effects (top) and $t$-distributed random effects (bottom).}
\end{figure}

\begin{figure}[htp]
\begin{center}
\begin{tabular}{c}
\includegraphics[height=90mm]{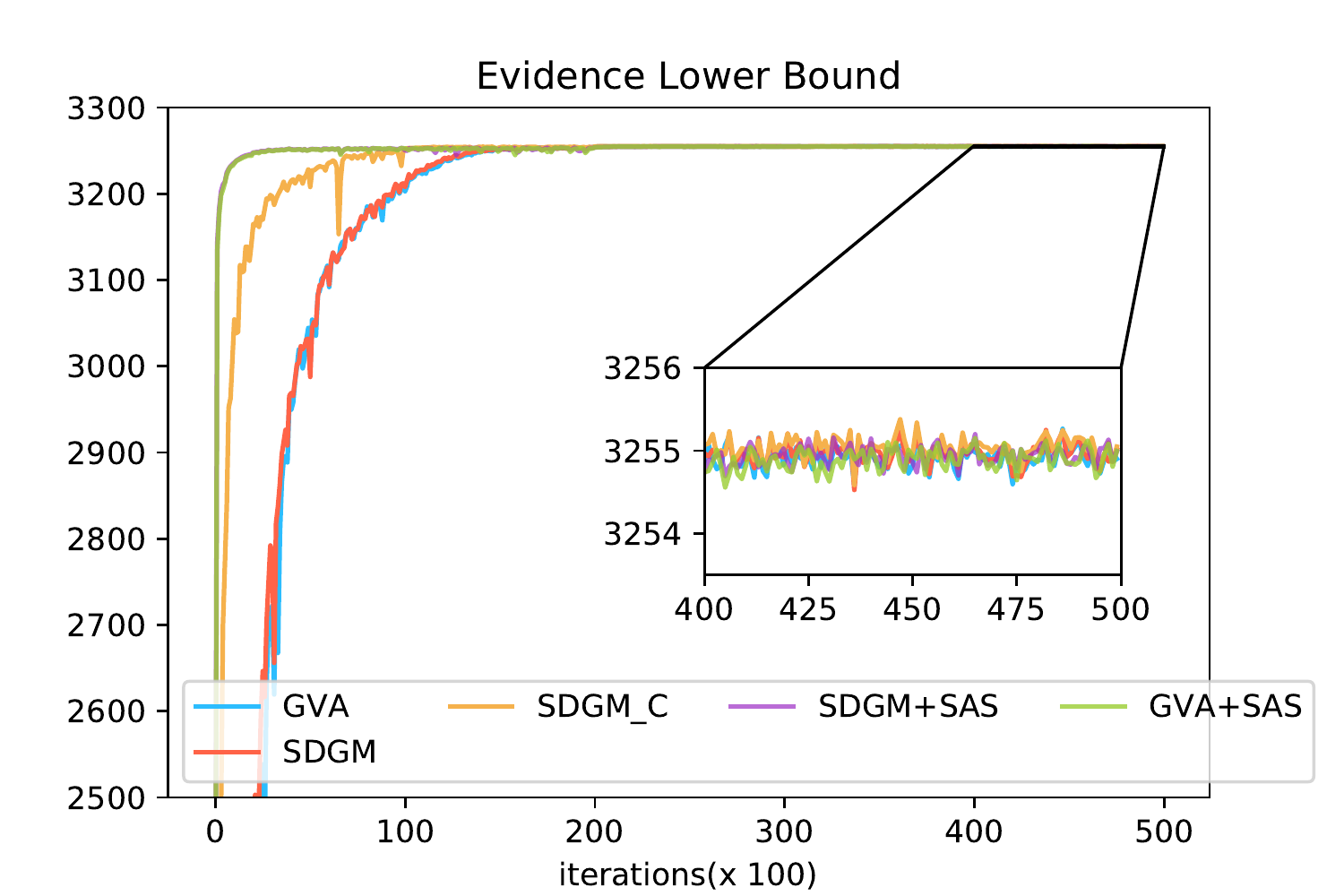} \\
\includegraphics[height=90mm]{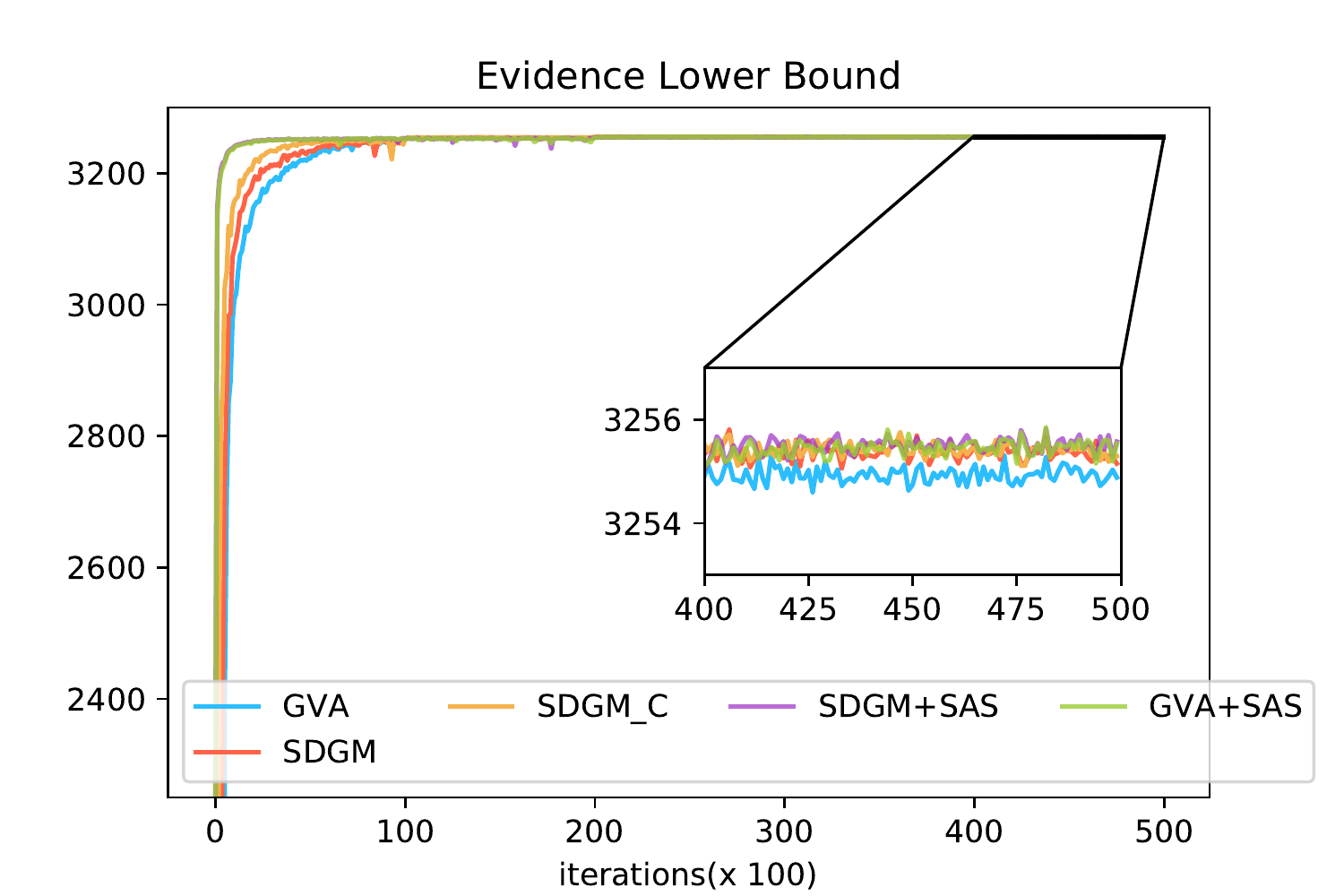}
\end{tabular}
\end{center}
\caption{\label{epilepsy-elbo}Monte Carlo estimate of ELBO versus iteration number for epilepsy data and normal random effects (top) and
$t$-distributed random effects (bottom).}
\end{figure}

Figure \ref{epilepsy-t} plots means, standard
deviations and skewness of variational methods versus
MCMC for the epilepsy example for $t$-distributed
random effects.  
Figures \ref{epilepsy-m} and \ref{epilepsy-elbo}  plot
the marginal posterior densities for the fixed effect
and variance parameters, and the ELBO plots versus iteration
number for the epillepsy example.

\begin{figure}[H]
\begin{center}
\begin{tabular}{c}
\includegraphics[height=50mm]{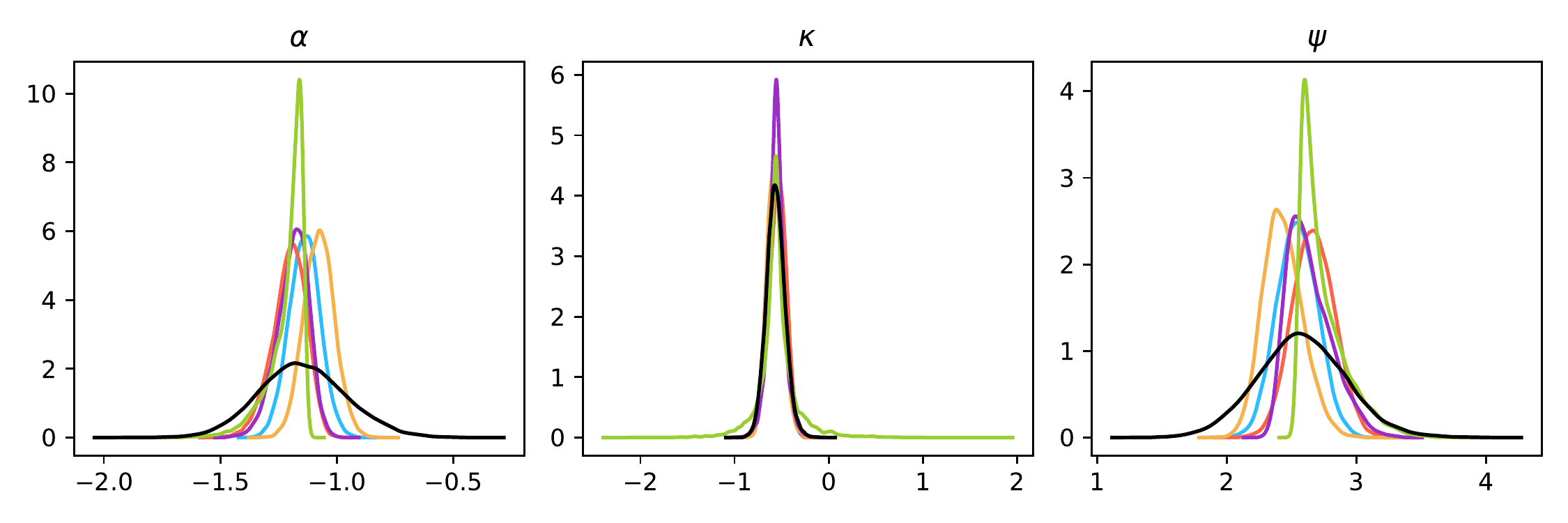} 
\end{tabular}
\end{center}
\caption{\label{NYSE-m}Comparison of marginal posterior densities
for $\alpha$, $\kappa$ and $\psi$ for the NYSE data.}
\end{figure}

Figure \ref{NYSE-m}  plots the marginal posterior densities
for the global parameters in the NYSE example.

The performance of all the SDGM and copula methods are
mostly similar for the random effects examples, but the Gaussian approximation
tends to perform poorly for estimating the  variance
parameters.  For the NYSE example, it is hard to discern
any improvement of the SDGM and copula methods compared to a
Gaussian approximation.

\end{document}